\documentclass[preprint,12pt]{elsarticle}

\usepackage{amsmath,amssymb,amsfonts}
\usepackage{algorithm}
\usepackage{algorithmic}
\usepackage[normalem]{ulem}
\usepackage{graphicx}
\usepackage{textcomp}
\usepackage{siunitx}
\usepackage{comment}
\usepackage{xcolor}
\usepackage{caption}
\usepackage{kotex}
\usepackage{subcaption}

\journal{arXiv}
\begin{document}

\begin{frontmatter}

%% Title, authors and addresses

%% use the tnoteref command within \title for footnotes;
%% use the tnotetext command for theassociated footnote;
%% use the fnref command within \author or \affiliation for footnotes;
%% use the fntext command for theassociated footnote;
%% use the corref command within \author for corresponding author footnotes;
%% use the cortext command for theassociated footnote;
%% use the ead command for the email address,
%% and the form \ead[url] for the home page:
%% \title{Title\tnoteref{label1}}
%% \tnotetext[label1]{}
%% \author{Name\corref{cor1}\fnref{label2}}
%% \ead{email address}
%% \ead[url]{home page}
%% \fntext[label2]{}
%% \cortext[cor1]{}
%% \affiliation{organization={},
%%             addressline={},
%%             city={},
%%             postcode={},
%%             state={},
%%             country={}}
%% \fntext[label3]{}

\title{Self-Consistent Numerical Framework for Multiscale Circuit--Plasma Coupling with Secondary Electron Emission
}

%% use optional labels to link authors explicitly to addresses:
%% \author[label1,label2]{}
%% \affiliation[label1]{organization={},
%%             addressline={},
%%             city={},
%%             postcode={},
%%             state={},
%%             country={}}
%%
%% \affiliation[label2]{organization={},
%%             addressline={},
%%             city={},
%%             postcode={},
%%             state={},
%%             country={}}
\author[postech]{Hongbin Kim}
\author[postech]{Soung Yong Yun}
\author[postech]{Jaeguk Lee}
\author[postech]{Dong-Yeop Na\corref{cor1}}

\affiliation[postech]{
  organization={Department of Electrical Engineering, Pohang University of Science and Technology (POSTECH)},
  addressline={77 Cheongam-ro, Nam-gu},
  city={Pohang},
  postcode={37673},
  state={Gyeongsangbuk-do},
  country={Republic of Korea}
}

\cortext[cor1]{Corresponding author}
\ead{dyna22@postech.ac.kr}

%\maketitle

%% Abstract

\begin{abstract}
Voltage breakdown in high-voltage pulsed vacuum systems arises from strongly nonlinear and multiscale interactions among external circuit dynamics, kinetic plasma evolution, and ion-induced secondary electron emission (SEE) at electrode surfaces. 
Although circuit--plasma co-simulation frameworks have been developed to couple lumped-element networks with particle-in-cell (PIC) plasma solvers, most existing approaches neglect energy-resolved SEE physics and its self-consistent feedback to both plasma and circuit, limiting predictive capability for breakdown initiation and transient surge dynamics.

In this work, we present a self-consistent numerical framework for multiscale circuit--plasma coupling that explicitly incorporates ion-energy-dependent SEE into the electrode boundary formulation of an electrostatic PIC solver. 
The emitted electron flux is directly included in the surface charge update, resulting in a modified Poisson boundary condition that couples plasma fields and circuit response within a unified formulation.
To address temporal multiscale effects, we derive two integration strategies: (i) a fully implicit strict coupling scheme that solves the plasma--circuit system monolithically at each time step, and (ii) a weak coupling scheme interpreted as a first-order operator-splitting approximation compatible with external SPICE-based circuit solvers, enabling a partitioned time-integration scheme with one-step lagged coupling between the circuit and plasma subsystems.
The weak formulation enables generalized embedding of arbitrary lumped-element networks while maintaining quantitative agreement with the strict reference solution.
The proposed framework is applied to voltage breakdown in a Tesla-transformer-driven vacuum capacitor with energetic ion injection.
Simulation results show that ion-induced SEE fundamentally alters surface charge evolution, triggering rapid voltage collapse and sustaining the experimentally observed near-zero-voltage plateau, whereas SEE-free models fail to reproduce these behaviors. 
Agreement between strict and weak coupling results confirms the robustness of the partitioned strategy.
The proposed method establishes a unified computational framework for predictive simulation of multiscale circuit--plasma interactions in high-power pulsed vacuum systems.
\end{abstract}

%%Graphical abstract
\begin{comment}
\begin{graphicalabstract}
%\includegraphics{grabs}
\end{graphicalabstract}
\end{comment}

\begin{comment}
%%Research highlights
\begin{highlights}
\item Research highlight 1
\item Research highlight 2
\end{highlights}
\end{comment}

%% Keywords

\end{frontmatter}

%% Add \usepackage{lineno} before \begin{document} and uncomment 
%% following line to enable line numbers
%% \linenumbers

%% main text
%%

%% Use \section commands to start a section

\section{Introduction} \label{sec:Introduction}
{\color{black}
High-voltage pulsed vacuum devices~\cite{pai1995introduction,mesyats2005pulsed,korovin2004pulsed,kim2016review}, including plasma-based ion injectors and particle-beam-driven sources~\cite{korovin2004pulsed,horioka2018progress,gurinovich2025explosive}, are typically operated with external driving circuits such as Tesla-transformer-based generators~\cite{white1948pulse,korovin1996high,balcerak2017compact,zhang2008experimental,su2009long} and pulsed power modulators~\cite{rossi2011advances,kuzmichev2003investigation}. 
In these systems, the electrical load is intrinsically dynamic rather than passive: the effective impedance of the vacuum gap evolves as high-energy plasma forms and transports charge, thereby modifying the external circuit response in real time. 
This strong bidirectional coupling between plasma kinetics and circuit dynamics can produce transient voltage collapse, surge currents, and breakdown phenomena. 
Understanding and mitigating such instabilities require predictive analysis of the coupled circuit--plasma response, demanding numerical frameworks capable of resolving circuit and plasma evolution in a fully self-consistent manner across disparate temporal scales.

As a representative example, we consider a specific experimental configuration of a Tesla-transformer-driven vacuum system, where rapid voltage collapse is observed under energetic ion injection.
The experimental configuration, as illustrated in Fig~\ref{fig:observation}, maintains a high vacuum of $10^{-6}$~Torr to ensure collisionless plasma transport. A titanium-coated arc source is employed to generate a controlled injection of Ti plasma. To verify the proper injection of the Ti ions, a Rogowski coil is installed between the arc source and ground. A copper substrate is positioned opposite the arc source and is connected to the load side of the Tesla transformer, as shown in Fig.~\ref{fig:Simplified Tesla transformer circuit}. As seen from the Tesla transformer, the substrate-arc source assembly initially behaves as a purely capacitive load in a vacuum. However, the introduction of Ti ions transforms this passive gap into a dynamic load, initiating a strong feedback loop between the evolving plasma and the external circuit. Fig. 1 illustrates the macroscopic result of this interaction: under identical circuit conditions, ion injection leads to an abrupt voltage drop followed by a sustained near-zero-voltage plateau. Our analysis suggests that this instability cannot be explained solely by primary ion transport. Instead, energetic ions impacting
electrode surfaces generate secondary electron emission (SEE), which initiates complex plasma dynamics that strongly modify surface convection currents and circuit-level charge evolution.

Various circuit--plasma co-simulation approaches have been developed to couple lumped-element networks with electrostatic or electromagnetic particle-in-cell (PIC) solvers~\cite{verboncoeur1993simultaneous,vahedi1995monte,verboncoeur2005particle,smithe2009external,liu2017circuit}. 
While successful in modeling space-charge-limited currents and plasma-induced impedance variation, most existing frameworks neglect or oversimplify ion-induced SEE at electrode surfaces when applied to voltage breakdown phenomena. 
As a result, conventional circuit--plasma models fail to reproduce the experimentally observed voltage collapse even under artificially enhanced ion injection, indicating that SEE-driven plasma--circuit feedback is not properly embedded in current coupling formulations.

\begin{figure}[t]
    \centering
    \includegraphics[width=1\linewidth]{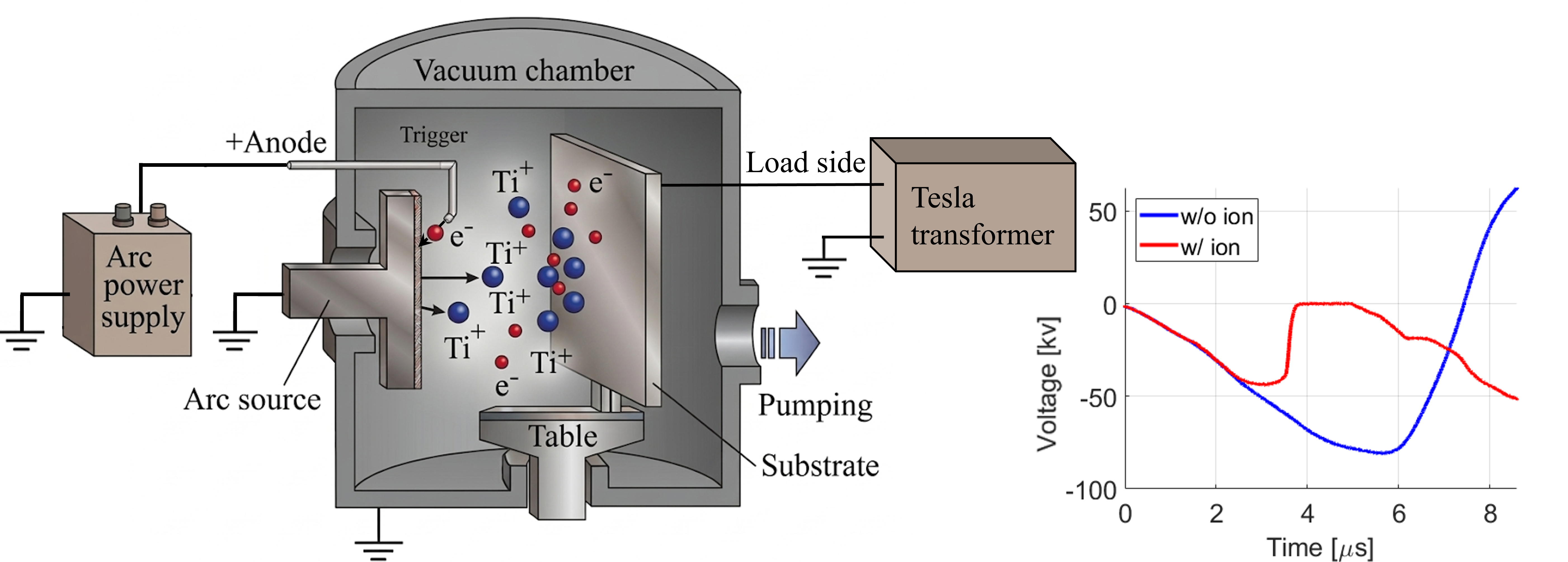}
    \caption{Schematic diagram of the cathodic arc-based \(\text{Ti}^+\) ion implantation and deposition system. (Created with the assistance of Nano Banana)}
    \label{fig:observation}
\end{figure}

Under high electric fields, ions accelerated across vacuum gaps can reach kinetic energies of several to tens of keV before impacting electrodes. 
The resulting SEE fundamentally alters sheath dynamics, enhances transient conduction currents, and injects additional charge into the external circuit, thereby reshaping voltage evolution at the system level. 
Although SEE-driven surface breakdown has been extensively studied in RF and high-power microwave systems~\cite{chang2011review,chang2014ultrafast,chang2015ion}, these studies typically prescribe the driving field and do not incorporate dynamic feedback with an external circuit.

Consequently, a gap remains between circuit--plasma frameworks that capture charge exchange and SEE-based breakdown models that neglect circuit-level feedback. 
In particular, existing formulations do not embed energy-resolved SEE into the electrode boundary condition in a manner that preserves self-consistent coupling between plasma kinetics and circuit equations.

In this work, we develop a self-consistent numerical framework for multiscale circuit--plasma coupling with ion-induced SEE. 
Energy-dependent SEE is incorporated directly into the electrode boundary formulation of an electrostatic PIC solver and coupled to a lumped-element circuit model. 
Specifically, we present two integration strategies: (i) a fully implicit strict coupling scheme that solves the plasma--circuit system monolithically at each time step, serving as a mathematically consistent reference formulation; and (ii) a weak partitioned scheme interpreted as a first-order operator-splitting approximation that enables time integration with lagged interfacial variables. 
While the strict formulation ensures rigorous self-consistency, the weak scheme facilitates interoperability with external circuit solvers and allows generalized embedding of arbitrary lumped-element networks. 
Quantitative agreement between the two approaches demonstrates the robustness of the proposed framework.
The proposed model is validated through comparison with measurements of voltage breakdown in a Tesla-transformer-driven vacuum capacitor with energetic ion injection. 
Simulations neglecting SEE fail to reproduce the observed voltage collapse, whereas the proposed SEE-enabled framework captures both breakdown thresholds and transient current evolution. 
These results establish SEE as a dominant mechanism in high-voltage pulsed breakdown and highlight the necessity of fully self-consistent circuit--plasma modeling.

The remainder of this paper is organized as follows.
Section~2 presents the numerical formulation.
Section~3 compares simulation results with measurements and analyzes breakdown mechanisms.
Section~4 concludes the paper.
\begin{comment}
The remainder of this paper is organized as follows.
Section~2 presents the formulation of the circuit-plasma-SEE co-simulation framework.
Section~3 describes the experimental setup and numerical implementation.
Section~4 compares simulation results with experimental measurements and analyzes the breakdown mechanism.
Finally, Section~5 summarizes the conclusions and outlines directions for future work.
\end{comment}
}

{\color{black}
\section{Self-Consistent Multiscale Circuit--Plasma Formulation with Secondary Electron Emission}

In this section, we present a self-consistent multiscale simulation framework 
that integrates a transient circuit solver, a kinetic plasma solver, 
and a SEE module. 
The objective is to capture the strongly coupled dynamics between 
(i) a resonant Tesla transformer circuit, 
(ii) a plasma formed inside a load capacitor, and 
(iii) surface charge modification induced by SEE.

The proposed framework is modular in structure. 
The transient circuit module describes the resonant energy exchange 
between inductive and capacitive elements of the Tesla transformer. 
The kinetic plasma module models the evolution of charged particles 
within the load capacitor using a one-dimensional electrostatic 
PIC method. 
The SEE module accounts for stochastic electron emission 
triggered by ion impact on the electrode surface 
using experimentally informed probability distributions.

Although each module describes a distinct physical subsystem, 
their interaction is intrinsically bidirectional. 
The circuit determines the voltage applied across the load capacitor, 
which drives ion acceleration and plasma formation. 
The evolving plasma redistributes charge in the inter-electrode gap, 
modifying the electric field and inducing surface charge variations on the electrodes. 
Furthermore, SEE alters the effective electrode charge 
through stochastic particle emission processes. 
These effects collectively influence the circuit current and voltage dynamics.

To resolve this tightly coupled multi-physics interaction, 
two integration strategies are introduced: 
a \emph{strict coupling mode}, in which circuit and plasma equations 
are assembled into a unified implicit formulation, 
and a \emph{weak coupling mode}, in which the two subsystems 
are advanced sequentially through exchange of interfacial variables. 
The strict mode ensures full self-consistency at each time step, 
while the weak mode provides improved flexibility and compatibility 
with general-purpose circuit solvers such as \texttt{PySpice}. 

The following subsections describe each module in detail, 
followed by the integration strategy and stability considerations.

\subsection{Transient circuit analysis module}

The transient response of the Tesla transformer circuit is modeled using two complementary approaches: 
(i) a manually derived solver based on Kirchhoff’s voltage and current laws (KVL/KCL), and 
(ii) a SPICE-based solver implemented via \texttt{PySpice}.

The manual solver provides full control over the governing equations and enables direct incorporation into a monolithic implicit formulation. 
This approach offers clear physical insight into energy exchange mechanisms and allows rigorous self-consistent coupling with the plasma solver. 
However, its scalability is limited, since the governing equations must be re-derived whenever the circuit topology is modified.

In contrast, the \texttt{PySpice}-based solver allows transient analysis of arbitrarily complex circuit configurations without manual reformulation. 
It offers high flexibility and practical convenience, particularly for dynamically evolving circuit topologies. 
Nevertheless, because the internal time integration is handled as a black-box process, direct embedding into a unified implicit circuit--plasma formulation is not straightforward.

\subsubsection{Manual solver option}
\label{sec:manual_solver}

\begin{figure}[t]
    \centering
    \includegraphics[width=1\linewidth]{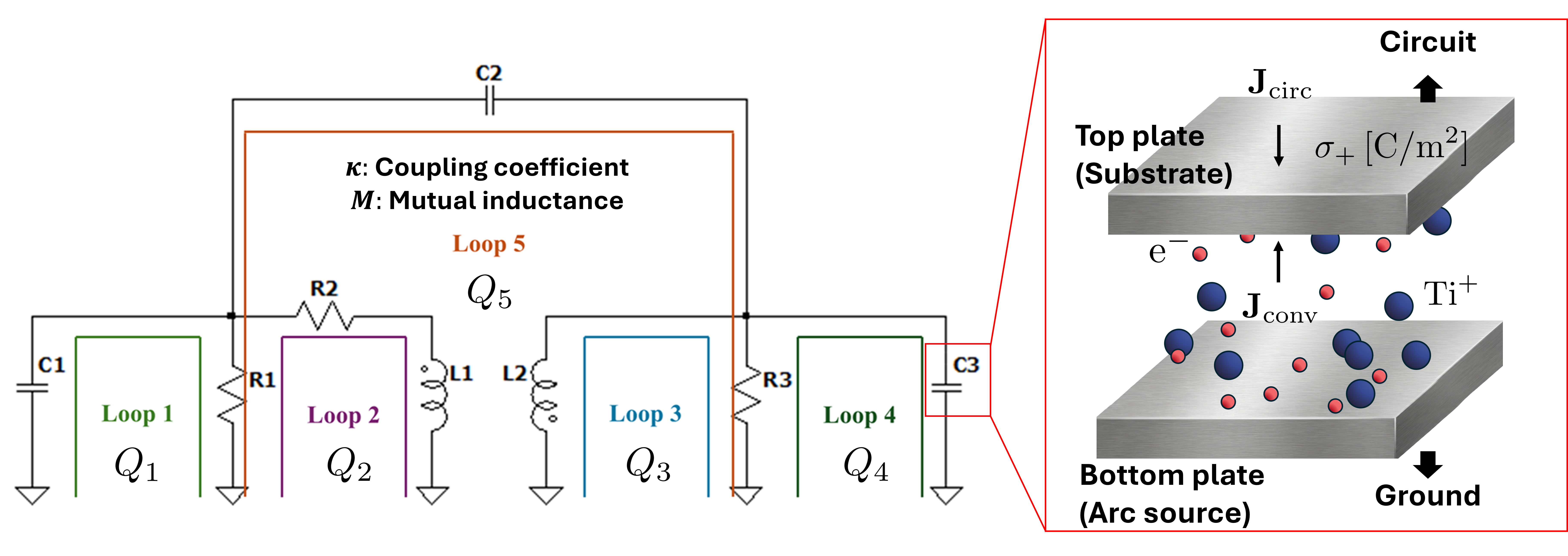}
    \caption{Simulation model comprising the simplified Tesla transformer circuit and the physical discharge domain.}
    \label{fig:Simplified Tesla transformer circuit}
\end{figure}

The simplified Tesla transformer circuit considered in this study is shown in Fig.~\ref{fig:Simplified Tesla transformer circuit}. 
The circuit consists of a source-side LC resonator $(C_1, L_1)$, a load-side LC resonator $(L_2, C_3)$, and a parasitic capacitance $C_2$. 
Initially, $C_1$ is charged by an external DC source, and at $t=0$ the source is disconnected, initiating transient oscillatory energy transfer between the magnetically coupled resonators.

To model the circuit dynamics explicitly, we formulate the governing equations using Kirchhoff’s voltage law (KVL) in terms of loop charges $Q_i$. 
Applying KVL to each loop yields

\begin{flalign}
\frac{Q_1}{C_1}
+
R_1\frac{d(Q_1-Q_2-Q_5)}{dt}
&= 0, 
\label{eq:loop_1}
\\
R_1\frac{d(-Q_1+Q_2+Q_5)}{dt}
+ R_2\frac{dQ_2}{dt}+ L_1\frac{d^2 Q_2}{dt^2}
+ M\frac{d^2 Q_3}{dt^2}
&= 0, 
\label{eq:loop_2}
\\
M\frac{d^2 Q_2}{dt^2}
+
L_2\frac{d^2 Q_3}{dt^2}
+
R_3\frac{d(Q_3-Q_4+Q_5)}{dt}
&= 0, 
\label{eq:loop_3}
\\
R_3\frac{d(-Q_3+Q_4-Q_5)}{dt}
+
\frac{Q_4}{C_3}
&= 0, 
\label{eq:loop_4}
\\
R_1\frac{d(-Q_1+Q_2+Q_5)}{dt}
+ \frac{Q_5}{C_2}
+ R_3\frac{d(Q_3-Q_4+Q_5)}{dt}
&= 0,
\label{eq:loop_5}
\end{flalign}
where $M$ denotes the mutual inductance between $L_1$ and $L_2$, and $Q_i$ represents the charge defined on the $i$-th loop.

For time discretization, we apply the second-order backward differentiation formula (BDF2) to the ODE system given in Eqs.~\eqref{eq:loop_1}--\eqref{eq:loop_5}. 
This procedure leads to a linear algebraic system at each time step of the form
\begin{equation}
\overline{\mathbf{T}} \cdot \mathbf{Q}^n = \mathbf{C},
\label{eq:matrix_form}
\end{equation}
where $\overline{\mathbf{T}}$ denotes the circuit coefficient matrix, the unknown state vector $\mathbf{Q}^n = [Q_1^n, Q_2^n, Q_3^n, Q_4^n, Q_5^n]^T$ contains the loop charges at time step $t_n = n\Delta t$, and 
$\mathbf{C}$ a RHS vector which collects all known contributions from previous time levels arising from the BDF2 discretization. 
The detailed derivation, including the explicit block-matrix expressions, is provided in \ref{App_circuit}.

At each time step, the circuit state is obtained by solving
\[
\mathbf{Q}^n
=
\overline{\mathbf{T}}^{-1} \cdot \mathbf{C}.
\]
Since $\overline{\mathbf{T}}$ depends only on circuit parameters and
$\Delta t$, it remains constant throughout the simulation and may
therefore be factorized once prior to time marching.
The initial conditions correspond to the configuration in which only capacitor $C_1$ is charged prior to switching, i.e.,
\begin{flalign}
Q_1(0)=Q, \qquad Q_2(0)=Q_3(0)=Q_4(0)=Q_5(0)=0.
\end{flalign}

\subsubsection{\texttt{PySpice} solver option}
\label{sec:pyspice_solver_main}
In addition to the manual circuit formulation described above, 
we also consider a SPICE-based circuit solver as an alternative option. 
Specifically, we employ \texttt{PySpice}, an open-source simulation framework 
built upon the \texttt{ngspice} engine.

In the present implementation, the circuit state is advanced 
in a time-marching manner consistent with the global time step $\Delta t$ 
of the coupled simulation. 
At each time step, the circuit variables from the previous step are 
imposed as initial conditions, and a transient analysis over 
the interval $[t_n, t_{n+1}]$ is performed. 
The resulting node voltages and inductor currents are then 
extracted and supplied to the plasma solver, enabling 
consistent interfacial data exchange within the integrated framework.

Because the internal time integration of \texttt{ngspice} is handled 
as a black-box process, the circuit equations cannot be assembled 
into a single monolithic implicit system together with the plasma equations. 
Consequently, this approach corresponds to a partitioned (weakly coupled) 
treatment of the circuit--plasma system. 
Nevertheless, it provides a robust and scalable solution strategy, 
particularly for complex circuit topologies where manual derivation 
of governing equations would be cumbersome.

For the circuit shown in Fig.~\ref{fig:Simplified Tesla transformer circuit}, 
the procedure for advancing the circuit state over one global time step 
$\Delta t$ is summarized in Algorithm~\ref{alg:circuit_solver}. 
The inputs are the circuit state variables at time step $n$, 
namely the node voltages $\mathbf{V}^{n}$ and inductor currents 
$\mathbf{I}^{n}$, together with $\Delta t$. 
The outputs are the updated circuit states 
$\mathbf{V}^{n+1}$ and $\mathbf{I}^{n+1}$.

\begin{algorithm}[t]
\caption{\texttt{PySpice}-based circuit solver}
\label{alg:circuit_solver}
\begin{algorithmic}[1]
    \REQUIRE Circuit state at time step $n$: $\mathbf{V}^{n}, \mathbf{I}^{n}, \Delta t$
    \ENSURE Circuit state at time step $n+1$: $\mathbf{V}^{n+1}, \mathbf{I}^{n+1}$

    \STATE \textbf{Initialize circuit:}
    \STATE \hspace{1em} Define topology ($C_1,C_2,C_3,L_1,L_2,R_1,R_2,R_3,M$).
    
    \STATE \textbf{Apply initial conditions:}
    \STATE \hspace{1em} Set capacitor voltages $\mathbf{V}_C^n = (V_{C1},V_{C2},V_{C3})$
    \STATE \hspace{1em} Set inductor currents $\mathbf{I}_L^n = (I_{L1},I_{L2})$
    
    \STATE \textbf{Perform transient analysis:}
    \STATE \hspace{1em} Run transient simulation for $\Delta t$ using
    \STATE \hspace{2em} \texttt{SIMULATOR.TRANSIENT(end\_time = $\Delta t$, use\_init = TRUE)}
    
    \STATE \textbf{Extract and update state:}
    \STATE \hspace{1em} $\mathbf{V}^{n+1} \leftarrow \texttt{SIMULATOR.ANALYSIS(Nodes)}$
    \STATE \hspace{1em} $\mathbf{I}^{n+1} \leftarrow \texttt{SIMULATOR.ANALYSIS(Inductors)}$
    
    \STATE \textbf{Return:} $\mathbf{V}^{n+1}, \mathbf{I}^{n+1}$
\end{algorithmic}
\end{algorithm}

\subsection{Kinetic plasma solver module}

In response to the transient voltage applied across the load capacitance, ion injection is initiated within the load capacitor $\mathrm{C3}$. 
The injected high-energy ions form a plasma between the two capacitor electrodes. 
The evolving charge distribution induces additional surface charges on the electrodes, dynamically modifying the voltage across the load capacitor through interaction with the external circuit.
To model this process, we employ a one-dimensional electrostatic particle-in-cell (ES-PIC) method.

\subsubsection{Field solver}

The electric potential satisfies the Poisson equation
\begin{equation}
\nabla^2 \phi = -\frac{\rho}{\epsilon_0},
\label{eq:Poisson_main}
\end{equation}
where $\phi$ is the electric potential and $\rho$ is the charge density obtained from particle deposition.
After spatial discretization based on the finite-difference method, the field equation is written in matrix form as
\begin{equation}
\overline{\mathbf{L}} \cdot \boldsymbol{\phi}^n
=
\boldsymbol{\rho}^n,
\label{eq:mt_Poisson}
\end{equation}
where $\overline{\mathbf{L}}$ denotes the Laplacian matrix and 
$\boldsymbol{\phi}^n$ is the vector of nodal potentials at time step $n$.
The detailed spatial discretization procedure and the explicit matrix 
representation of the resulting linear system are provided in 
\ref{App_plasma}.
It should be noted that the lower electrode is grounded, i.e.,
\begin{equation}
\phi_N^n = 0,
\end{equation}
where $N$ denotes the index of the last grid node.
In contrast, the upper electrode is not prescribed directly; instead, it is 
coupled self-consistently to the external circuit through Gauss’s law. 
This leads to the following discrete boundary condition:
\begin{equation}
-\phi_0^n + \phi_1^n
=
-\frac{\Delta x^2}{\epsilon_0}
\left(
\frac{\sigma_+^n}{\Delta x}
+
\frac{\rho_0^n}{2}
\right),
\label{eq:Poisson_BC}
\end{equation}
which establishes a direct relationship between the electrode potential 
$\phi_0^n$ and the surface charge density $\sigma_+^n$ supplied by the external circuit.

\subsubsection{Particle solver}

Once the electric field is obtained from the potential, particle motion is advanced using a standard leapfrog scheme:
\begin{flalign}
v^{n+\frac{1}{2}}_p &= v^{n-\frac{1}{2}}_p
+ \frac{q_p \Delta t}{m_p} E_p^n,
\label{eq:vel_update_main}
\\
x_p^{n+1} &= x_p^n
+ \Delta t \, v_p^{n+\frac{1}{2}},
\label{eq:pos_update_main}
\end{flalign}
where $p$ denotes the superparticle index.
Note that the magnetic rotation is neglected since an electrostatic approximation is assumed.
The electric field at particle positions is obtained via linear interpolation from grid-based field values. 
Details of the interpolation scheme and charge deposition are provided in \ref{App_interpolation_deposition}.

It should be noted that the circuit and plasma modules are coupled implicitly through the surface charge $\sigma_+^n$ and the electrode potential $\phi_0^n$. 
Since these quantities serve as inputs to each other at the same time step, the circuit and plasma equations must be solved in a self-consistent manner within each time step.

\subsection{Secondary Electron Emission Module}

\begin{figure}[t]
    \centering
    \includegraphics[width=0.8\linewidth]{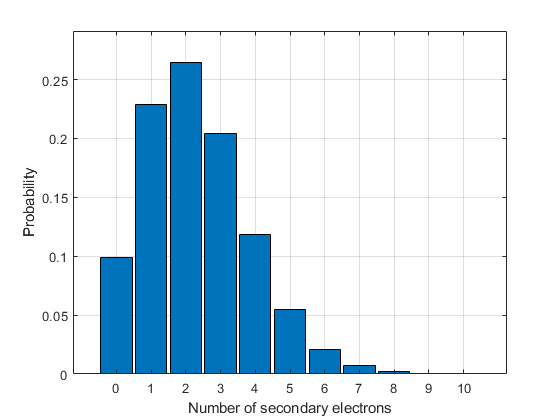}
    \caption{Probability mass function of emitted secondary electrons for \SI{40}{\kilo\electronvolt} incident ions.}
    \label{fig:Poisson}
\end{figure}

To reproduce realistic operating conditions, the SEE process is incorporated into the PIC simulation using a Monte Carlo approach based on probability distributions fitted to available experimental data.
Because the target experiment involves $\mathrm{Ti}^+$ ion injection into a Cu-based load capacitor, the SEE model is constructed to represent $\mathrm{Ti}^+$ ion impact on a Cu surface.

The modeling focuses on incident ion energies in the range of 0--\SI{80}{\kilo\electronvolt}, which covers the dominant ion energy distribution observed in the present pulsed system.
The following quantities are modeled probabilistically:
\begin{enumerate}
    \item the secondary electron yield (SEY) per ion impact as a function of incident ion energy,
    \item the kinetic energy distribution of emitted secondary electrons, and
    \item the emission angle distribution.
\end{enumerate}

For modeling simplicity in the present implementation, these quantities are assumed statistically independent.
To date, no experimental SEE measurements have been reported for $\mathrm{Ti}^+$ ions incident on Cu within the relevant energy range.
Most available data correspond to noble gas ions.
Therefore, experimental data for $\mathrm{Ar}^+$ ions are adopted as a proxy, since the ion mass and nuclear stopping characteristics are comparable to those of $\mathrm{Ti}^+$ in the considered energy regime.
Although this substitution does not capture detailed electronic structure effects, it provides a physically reasonable approximation for the ion-impact-induced emission behavior.

%-------------------------------------------------------
\subsubsection{Secondary electron yield modeling}

In the PIC simulations, the number of emitted secondary electrons per ion impact is sampled from a Poisson distribution (Fig.~\ref{fig:Poisson}), whose mean value is given by the experimentally reported SEY as a function of incident ion energy.

The energy dependence of the mean SEY is obtained by fitting a fifth-order polynomial to experimental data reported by Holm\'{e}n \textit{et al}.~\cite{Holmen1981} and Zalm and Beckers~\cite{Zalm1985} using a least-squares method for incident energies below \SI{80}{\kilo\electronvolt}, as shown in Fig.~\ref{fig:SE_yield}.
The fitted polynomial function is denoted by $f_{\mathrm{poly}}(E_{\mathrm{ion}})$.

\begin{figure}[t]
    \centering
    \includegraphics[width=0.8\linewidth]{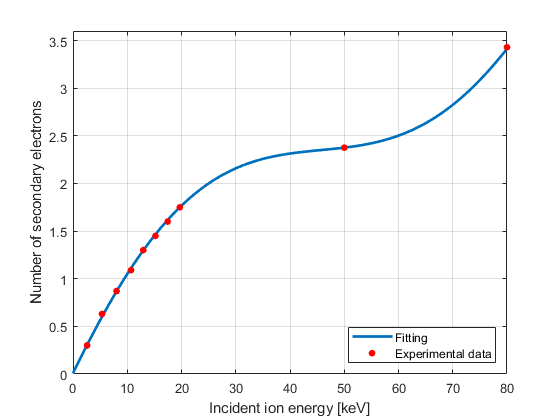}
    \caption{Energy dependence of the mean SEY. Experimental data are taken from~\cite{Holmen1981, Zalm1985}.}
    \label{fig:SE_yield}
\end{figure}

%-------------------------------------------------------
\subsubsection{Energy distribution modeling}

The energy distribution of emitted secondary electrons is modeled based on experimental spectra reported by Ruano and Ferr\'{o}n~\cite{Ruano2008}.
The published data are provided in the form of an energy-weighted spectrum, $N(E)\cdot E$, for incident ion energies up to \SI{5}{\kilo\electronvolt} (Fig.~\ref{fig:e_dis_ex}).
Although this representation cannot be directly interpreted as a probability density function suitable for Monte Carlo sampling, it reveals an important physical feature:
independently of the incident ion energy, the emitted secondary electrons predominantly occupy the low-energy inelastic regime, particularly below \SI{20}{\electronvolt}.

To capture this experimentally observed trend while maintaining numerical robustness, a lognormal distribution is adopted to model the probability density of secondary electron energy for all incident ion energies (Fig.~\ref{fig:e_dis}).
The parameters $(\mu,\sigma)$ are selected to reproduce the experimentally observed peak and width of the emission spectrum.

\begin{figure}
    \centering
    \includegraphics[width=0.8\linewidth]{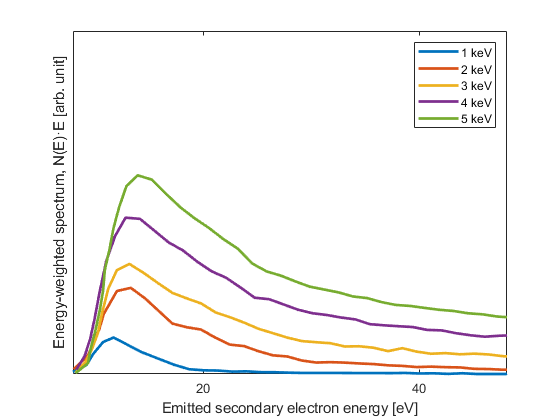}
    \caption{Energy-weighted secondary electron energy spectra adapted from~\cite{Ruano2008}. Here, $N(E)$ denotes the number of emitted secondary electrons with kinetic energy $E$.}
    \label{fig:e_dis_ex}
\end{figure}

\begin{figure}
    \centering
    \includegraphics[width=0.8\linewidth]{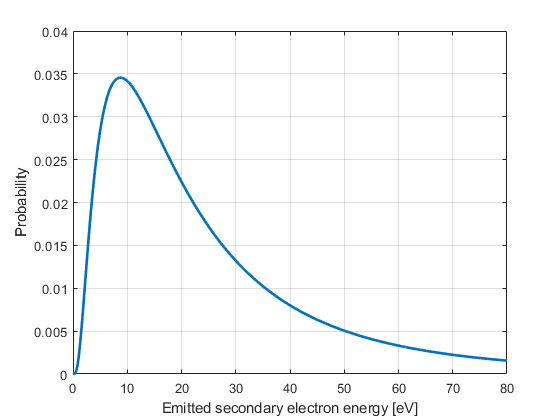}
    \caption{Adopted lognormal probability density function for emitted secondary electron energy.}
    \label{fig:e_dis}
\end{figure}

%-------------------------------------------------------
\subsubsection{Monte Carlo SEE sampling algorithm}

The SEE process is implemented through the following Monte Carlo sampling procedure.

\begin{algorithm}[H]
\caption{SEE Sampling Procedure (1D)}
\label{alg:see_sampling_1d}
\begin{algorithmic}[1]
    \REQUIRE Ion velocity $v_{x,\mathrm{ion}}$, ion mass $m_{\mathrm{ion}}$, electron mass $m_{\mathrm{e}}$
    \ENSURE Number of emitted electrons $n_{\mathrm{SE}}$ and emitted velocities $v_{x,\mathrm{SE}}$

    \STATE \textbf{Compute incident ion energy:}
    $
    E_{\mathrm{ion}} \gets \frac{1}{2} m_{\mathrm{ion}} v_{x,\mathrm{ion}}^2
    $

    \STATE \textbf{Evaluate mean SEY:}
    $
    \bar{n}_{\mathrm{SE}} \gets f_{\mathrm{poly}}(E_{\mathrm{ion}})
    $,~$\bar{n}_{\mathrm{SE}} \gets \max(\bar{n}_{\mathrm{SE}}, 0)$

    \STATE \textbf{Sample number of emitted electrons:}
    $
    n_{\mathrm{SE}} \sim \mathrm{Poisson}(\lambda = \bar{n}_{\mathrm{SE}})
    $

    \FOR{$k = 1$ to $n_{\mathrm{SE}}$}
        \STATE Sample secondary electron energy:
        $
        E_{\mathrm{SE}} \sim \mathrm{Lognormal}(\mu, \sigma)
        $
        \STATE Convert to velocity magnitude:
        $
        v_{\mathrm{SE}} \gets \sqrt{\frac{2 E_{\mathrm{SE}}}{m_{\mathrm{e}}}}
        $
        \STATE Assign outward emission direction (1D normal emission)
    \ENDFOR

    \STATE \textbf{Return:} $n_{\mathrm{SE}}, v_{x,\mathrm{SE}}$
\end{algorithmic}
\end{algorithm}

\subsection{Integration of circuit, plasma, and SEE modules}
\label{sec:Integration}
We now describe the self-consistent integration of the one-dimensional PIC
algorithm, the transient circuit solver, and the SEE module.
These subsystems are dynamically coupled through two interfacial quantities:
the electrode potential and the net convection current at the electrode surface.
The convection current represents the net charge exchange between the plasma
region and the electrode, including contributions from incident ions,
impinging electrons, and the effective positive surface charge that develops
as a consequence of SEE.
Two distinct integration strategies are considered, referred to as the
\emph{strict} and \emph{weak} coupling modes.
The difference between the two lies in how the circuit-side charge response
is incorporated into the Poisson boundary condition at the electrode.

\begin{figure}[t]
    \centering
    \includegraphics[width=0.8\linewidth]{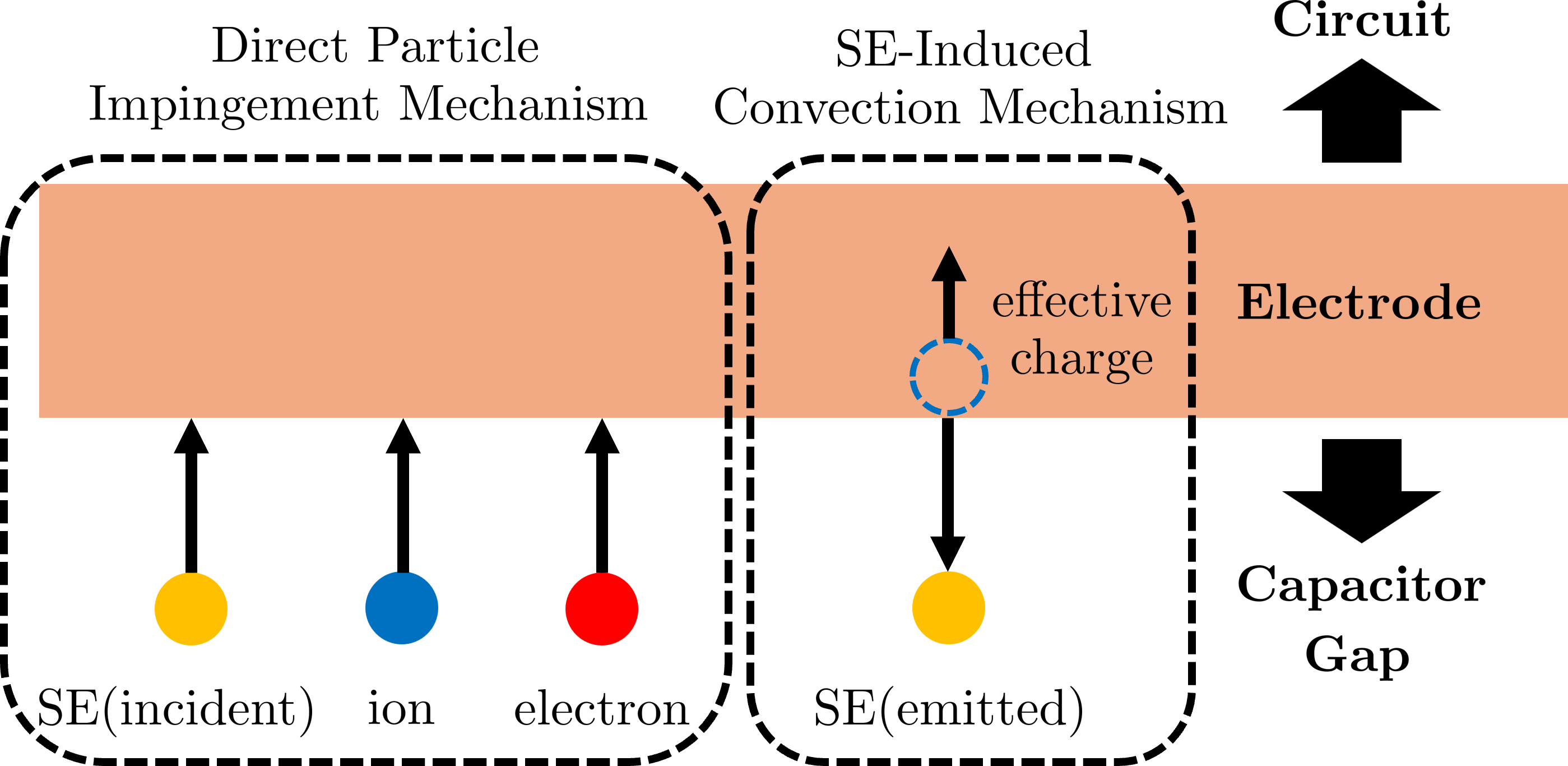}
    \caption{Two mechanisms contributing to the convection current in the present simulation:
    (i) direct particle impingement and
    (ii) secondary-electron-induced charge removal.}
    \label{fig:Convection_charge}
\end{figure}
\subsubsection{Convection current modeling}

The convection current is defined as the current component associated with
the direct transport of free charges across the plasma--electrode interface.
Unlike displacement current, which arises from time variation of the electric field,
the convection current corresponds to the physical arrival or departure of
charged particles at the electrode surface.

There are two primary mechanisms governing this interaction.

First, as illustrated by the direct particle impingement mechanism in
Fig.~\ref{fig:Convection_charge}, charged particles present in the inter-electrode
gap—including ions, background electrons, and previously emitted secondary electrons—
directly modify the electrode surface charge when they strike the surface.
In this mechanism, the surface charge variation arises purely from the
transport of charged particles across the interface.
The resulting rate of surface charge change corresponds to the convection
current in its conventional sense and has been explicitly accounted for
in earlier circuit--plasma co-simulation frameworks
\cite{verboncoeur1993simultaneous, smithe2009external}.

Second, as illustrated by the secondary-electron-induced mechanism in
Fig.~\ref{fig:Convection_charge}, SEE also modifies the electrode surface charge.
When a secondary electron is emitted, negative charge is removed from the
electrode and injected into the plasma.
This removal of electrons produces a transient electron deficit at the surface,
which is electrically equivalent to the formation of an effective positive
surface charge~\cite{Hollmann2006}.
Since this charge imbalance results from the direct transport of electrons
from the electrode into the plasma region,
the secondary-electron-induced charge deficit is treated in the present work
as part of the convection charge.

Accordingly, the total convection charge accumulated during a time step
can be expressed as
\begin{equation}
Q_{\mathrm{conv}}
=
Q_{\mathrm{ion}}
+
Q_{\mathrm{elec}}
+
Q_{\mathrm{SE}}^{\mathrm{inc}}
-
Q_{\mathrm{SE}}^{\mathrm{em}},
\end{equation}
where the individual terms represent the charges associated with incident ions,
incident plasma electrons, incident secondary electrons returning to the surface,
and emitted secondary electrons, respectively.

In the discrete time-marching simulation,
the convection charge at time step $n$ is evaluated by directly counting
the net charge transported across the electrode boundary during
the interval $[t_{n-1}, t_n]$.
It is therefore computed as
\begin{equation}
Q_{\mathrm{conv}}^{n}
=
\sum_{p \in \mathcal{P}_{n}^{\mathrm{ion}}} q_p
+
\sum_{p \in \mathcal{P}_{n}^{\mathrm{elec}}} q_p
+
\sum_{p \in \mathcal{P}_{n}^{\mathrm{SE,inc}}} q_p
-
\sum_{p \in \mathcal{P}_{n}^{\mathrm{SE,em}}} q_p ,
\end{equation}
where $q_p$ denotes the electric charge carried by particle $p$, and
\(\mathcal{P}_{n}^{\mathrm{ion}}\),
\(\mathcal{P}_{n}^{\mathrm{elec}}\),
\(\mathcal{P}_{n}^{\mathrm{SE,inc}}\), and
\(\mathcal{P}_{n}^{\mathrm{SE,em}}\)
represent the sets of incident ions, incident plasma electrons,
incident secondary electrons, and emitted secondary electrons
counted at time step $n$, respectively.

It should be emphasized that this definition ensures strict charge conservation
at the plasma--electrode interface.
The convection charge $Q_{\mathrm{conv}}^{n}$ enters directly into the
surface charge update equation in the coupled circuit--plasma formulation,
thereby providing the physical bridge between particle kinetics and
circuit-level dynamics.

\subsubsection{Strict coupling mode}
\label{sec:strict coupling mode}

In the strict coupling mode, the circuit and plasma subsystems are assembled
into a single implicitly coupled formulation through the electrode surface
charge density.
The total surface charge density on the top electrode, denoted by
$\sigma_{+}^{n}$, serves as the primary coupling variable that bridges
the circuit dynamics and the plasma kinetics.

Based on charge conservation at the electrode surface,
the surface charge density at time step $n$ can be written as
\begin{equation}
\sigma_+^n
=
\sigma_+^{n-1}
+
\frac{Q_{conv}^n+Q_{circ}^n - Q_{circ}^{n-1}}{A},
\label{eq:top_plate_charge}
\end{equation}
where $A$ denotes the electrode area.
Here, $Q_{conv}^n$ represents the net convection charge delivered
to the electrode from the plasma region during the time interval $\Delta t$.
This term includes contributions from incident ions,
impinging plasma electrons,
incident secondary electrons,
and the effective positive surface charge created by secondary-electron emission.
The term $Q_{circ}^n - Q_{circ}^{n-1}$ denotes the net charge supplied to
(or removed from) the electrode by the external circuit during the same time interval.
In the circuit configuration shown in
Fig.~\ref{fig:Simplified Tesla transformer circuit},
the circuit-side charge associated with the load capacitor is
identified as
\begin{equation}
Q_{circ} \equiv Q_4 .
\end{equation}
In the standalone circuit solver,
$Q_4^n$ is obtained from the discretized circuit system
given in Eqs.~\eqref{eq:matrix_form}.
However, in the coupled simulation,
the physical electrode charge is represented by the surface charge density
$\sigma_+^n$ rather than by the lumped variable $Q_4^n$.
Consequently, the voltage across the load capacitor $C_3$
can no longer be expressed as $Q_4/C_3$.
Instead, it must be determined consistently from the electrode potential
$\phi_0^n$ obtained from the Poisson solver.
To enforce this consistency, the loop equation associated with $C_3$,
originally given by Eq.~\eqref{eq:loop_4},
is modified as
\begin{equation}
R_3 \frac{d(Q_4 - Q_3)}{dt} + V_{C3} = 0 ,
\label{eq:loop_4_mod}
\end{equation}
where $V_{C3}$ is identified with the electrode potential
$\phi_0^n$.
Accordingly, the fourth block row of the circuit matrix
$\overline{\mathbf{T}}$
and the corresponding right-hand-side vector $\mathbf{C}$
are replaced by
$\overline{\mathbf{T}}'_4$ and $\mathbf{C}'$,
defined as
\begin{align}
\overline{\mathbf{T}}'_4 &=
\begin{bmatrix}
\dfrac{3R_3}{2\Delta t} & -\dfrac{3R_3}{2\Delta t}\\[1.2ex]
-\dfrac{3R_3}{2\Delta t} &
\dfrac{3R_1}{2\Delta t}
+\dfrac{3R_3}{2\Delta t}
+\dfrac{1}{C_2}
\end{bmatrix},
\label{eq:T4_matrix_full}
\\
\mathbf{C}' &=
\left[
\begin{matrix}
c_1 & c_2 & c_3 & c_4 - \phi_0^n & c_5
\end{matrix}
\right]^{n}.
\label{eq:circuit_matrix_mod}
\end{align}
Solving the modified circuit system yields an explicit relation
between the circuit charge $Q_4^n$ and the electrode potential $\phi_0^n$:
\begin{equation}
Q_4^n
=
K
-
\left[{\overline{\mathbf{T}}'}^{-1}\right]_{4,4}\,\phi_0^n ,
\label{eq:Q4_phi_relation}
\end{equation}
where
\begin{equation}
K
=
\sum_{i=1}^{5}
\left[{\overline{\mathbf{T}}'}^{-1}\right]_{4,i}\, c_i .
\label{eq:K_def}
\end{equation}
Substituting Eq.~\eqref{eq:Q4_phi_relation} into
Eq.~\eqref{eq:top_plate_charge} gives
\begin{equation}
\sigma_+^n
=
\sigma_+^{n-1}
+
\frac{
Q_{conv}^n
+
K
-
\left[{\overline{\mathbf{T}}'}^{-1}\right]_{4,4}\,\phi_0^n
-
Q_4^{n-1}
}{A}.
\label{eq:sigma_update}
\end{equation}
Finally, inserting Eq.~\eqref{eq:sigma_update}
into the electrode boundary condition of the Poisson equation
[Eq.~\eqref{eq:Poisson_BC}]
yields
\begin{align}
\left(
1 +
\frac{
\left[{\overline{\mathbf{T}}'}^{-1}\right]_{4,4}
\,\Delta x
}{\epsilon_0 A}
\right)\phi_0^n
-
\phi_1^n
=
\frac{\Delta x^2}{\epsilon_0}
\left[
\frac{1}{\Delta x}
\left(
\sigma_+^{\,n-1}
+
\frac{
Q_{conv}^n
+K
-Q_4^{n-1}
}{A}
\right)
+
\frac{\rho_0^n}{2}
\right].
\label{eq:Poisson_BC2}
\end{align}
By replacing the first row of the discrete Poisson matrix in
Eq.~\eqref{eq:mt_Poisson} with Eq.~\eqref{eq:Poisson_BC2},
the electric potential $\boldsymbol{\phi}^n$,
including the electrode potential $\phi_0^n$,
is obtained from a single linear solve that incorporates
both plasma quantities $(\rho^n, Q_{conv}^n)$
and circuit quantities from the previous time level
$(Q_{circ}^{n-1}, \sigma_+^{n-1})$.
The electric field $\mathbf{E}^n$ is then computed using
Eq.~\eqref{eq:dt_E-field} in~\ref{App_plasma},
and particle positions and velocities are advanced using the PIC pusher,
yielding $\rho^{n+1}$ and $Q_{conv}^{n+1}$.
The circuit and plasma subsystems are thus advanced in a fully
self-consistent manner within the same time level.
This procedure constitutes the \emph{strict coupling mode},
in which the circuit-side charge response is explicitly embedded
into the Poisson boundary condition,
resulting in a monolithic implicit formulation
of the coupled circuit--plasma system.
The overall workflow of the strict coupling mode
is illustrated in Fig.~\ref{fig:strict_mode_flowchart},
and the definitions of all variables are summarized in
Table~\ref{tab:variables}.
\begin{figure}[t]
\centering 
\includegraphics[width=0.8\linewidth]{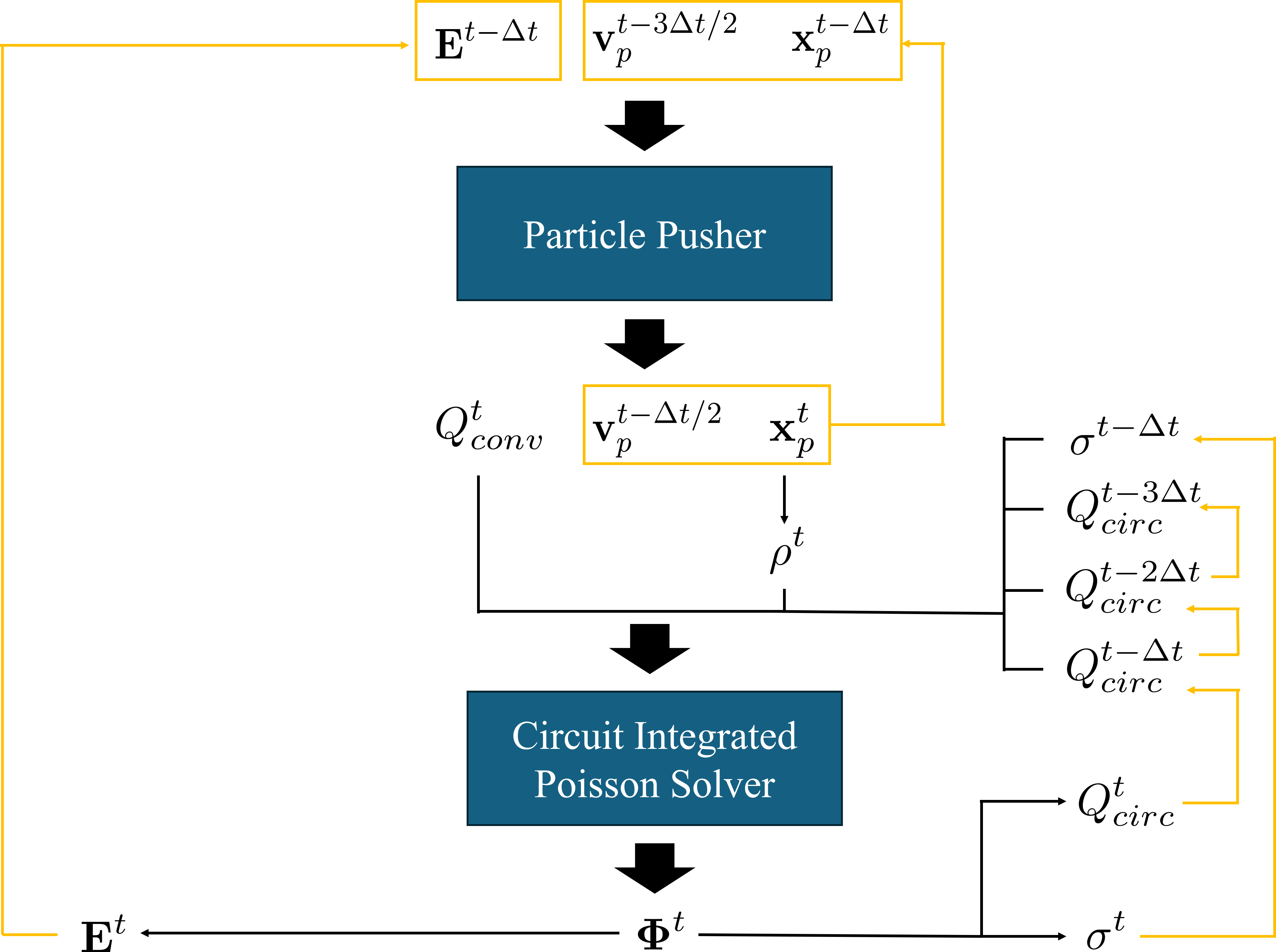} 
\caption{Flowchart of the strict coupling mode simulation.} 
\label{fig:strict_mode_flowchart} 
\end{figure}

\begin{table}[t]
\centering
\footnotesize
\caption{Definition of key variables}
\label{tab:variables}
\begin{tabular}{|c|l|}
\hline
\textbf{Variable} & \textbf{Physical Meaning} \\
\hline
$\mathbf{E}$ & Electric field at half-grid points \\
$\mathbf{v}_p$ & Individual particle velocities \\
$\mathbf{x}_p$ & Individual particle positions \\
$Q_{conv}$ & Convection charge \\
$\rho$ & Charge density between electrodes \\
$\sigma$ & Charge density of electrode surface\\
$Q_{circ}$ & Circuit charge \\
$\mathbf{\Phi}$ & Potential at nodes \\
\hline
\end{tabular}
\end{table}

\subsubsection{Weak coupling mode}
\label{sec:weak_coupling}

In practical implementations, general-purpose circuit simulators such as
\texttt{ngspice} (accessed via \texttt{PySpice})
employ implicit time-integration schemes,
including the trapezoidal rule and Gear methods.
Nonlinear algebraic systems arising at each time step are solved
internally using Newton--Raphson iterations.
However, the internal Jacobian matrices and time-discretized governing
equations are not directly accessible to the user.
As a consequence, the circuit subsystem cannot be assembled together
with the plasma equations into a single monolithic implicit system.
When integrated with the PIC solver,
the \texttt{PySpice}-based circuit solution therefore leads to
a partitioned coupling strategy,
referred to here as the \emph{weak coupling mode}.

While the strict coupling mode embeds the circuit-side charge
$Q_{circ}^n$ explicitly as a function of the electrode potential $\phi_0^n$,
the weak coupling mode advances the circuit and plasma subsystems
sequentially within each global time step.
Specifically, the circuit equations in
Eq.~\eqref{eq:matrix_form}
are first solved independently,
yielding the circuit-side charge ${Q_4^n}^{*}$ where the superscript $*$ denotes that the quantity
is obtained solely from the circuit simulation,
without incorporating the updated plasma state at time level $n$.
This independently computed charge
is then substituted into the surface-charge update equation
[Eq.~\eqref{eq:top_plate_charge}],
by identifying $Q_{circ}^n \equiv {Q_4^n}^{*}$.
Since most circuit simulators provide time-resolved node voltages,
the circuit-side charge can be obtained directly from the
load-capacitor voltage using ${Q_4^n}^{*} = C_3 \, V_{C3}^n$.
Substituting this expression into the electrode boundary condition
of the Poisson equation yields
\begin{align}
-\phi_0^n + \phi_1^n
=
-\frac{\Delta x^2}{\epsilon_0}
\left[
\frac{1}{\Delta x}
\left(
\sigma_+^{n-1}
+
\frac{
Q_{conv}^n
+
{Q_4^n}^{*}
-
Q_4^{n-1}
}{A}
\right)
+
\frac{\rho_0^n}{2}
\right].
\label{eq:Poisson_BC3}
\end{align}
The Poisson equation is then solved to obtain $\phi_0^n$ and the
electric field $\mathbf{E}^n$,
after which the particle positions and velocities are advanced,
yielding $\rho^{n+1}$ and $Q_{conv}^{n+1}$.
At the next time step,
the circuit state is updated using $Q_4^n = C_3 \, \phi_0^n$, which ensures consistency of the initial condition for the subsequent
circuit solve.

In contrast to the strict coupling formulation,
the electrode potential $\phi_0^n$
does not appear implicitly inside the circuit system matrix.
Instead, the circuit response and plasma response are evaluated
sequentially, resulting in a one-step time lag between the two subsystems.
From a numerical-analysis perspective,
the weak coupling mode can therefore be interpreted as
a first-order operator-splitting approximation of the fully implicit
strict coupling scheme.
Although this approach sacrifices the exact monolithic
self-consistency of the strict formulation,
it offers substantial practical advantages:
(i) arbitrary circuit topologies can be incorporated without
re-deriving governing equations,
(ii) robust and well-validated circuit solvers can be utilized directly,
and (iii) extensibility to large or nonlinear circuit configurations
is significantly improved.
The overall data flow of the weak coupling mode
is illustrated in Fig.~\ref{fig:weak mode flowchart}.
The weak formulation remains accurate
provided that the global time step satisfies
the charge-variation constraint discussed in~\ref{App_time_segment}.
\begin{figure}[t]
    \centering
    \includegraphics[width=0.8\linewidth]{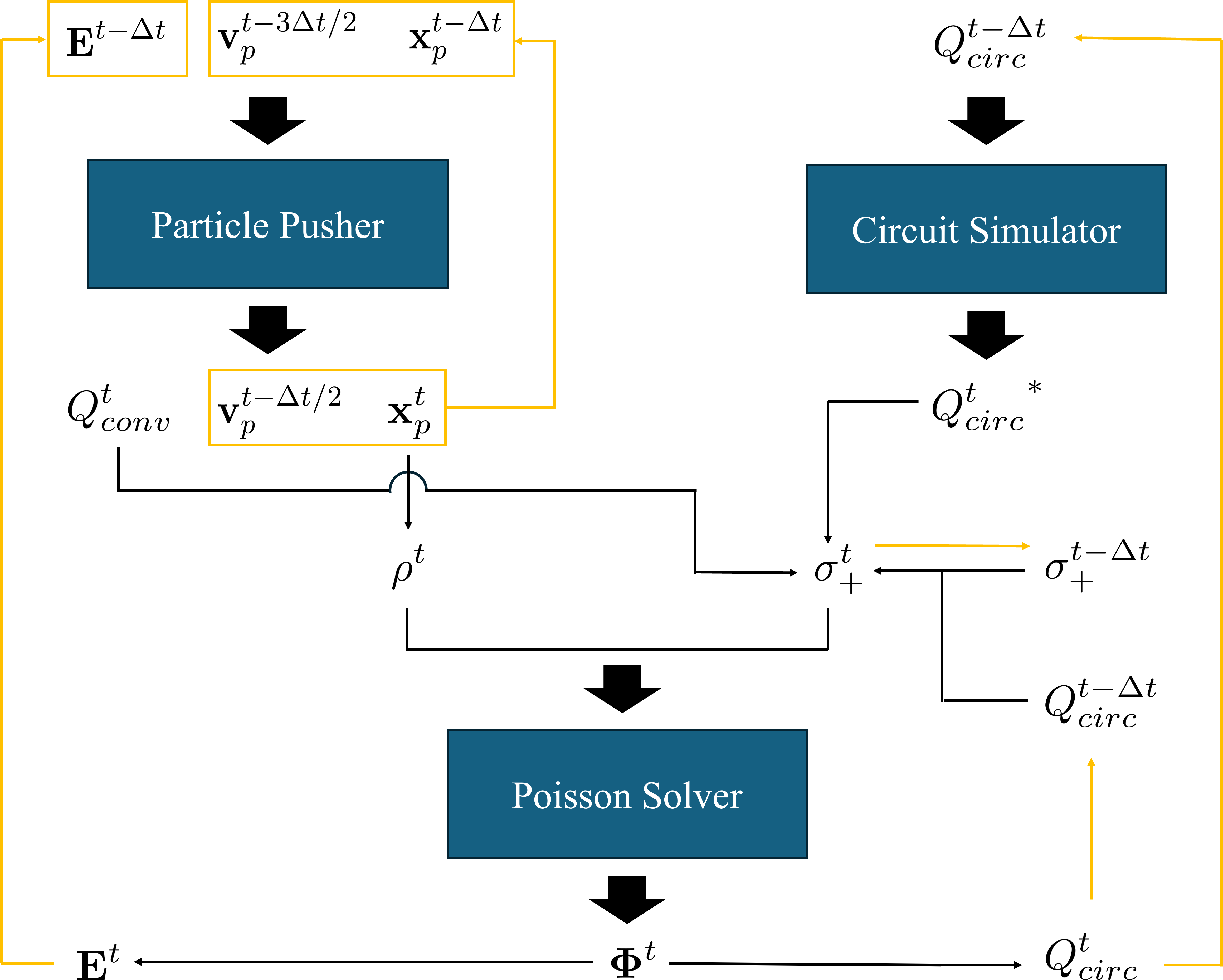}
    \caption{Flowchart of the weak coupling mode simulation.}
    \label{fig:weak mode flowchart}
\end{figure}

\section{Numerical Results and Discussion}
\label{sec:results}
We validate the proposed algorithm using a simple RC circuit in \ref{App_validation}.
In this section, we present numerical results demonstrating the performance of the proposed circuit--plasma interaction algorithm for modeling voltage breakdown in a Tesla transformer with ion source injection.

{\color{black}
\subsection{Tesla transformer without ion source injection}
\label{sec:result_1_w/o_ion}

We first verify the accuracy of the circuit solver prior to activating the ion-injection plasma module. 
The objective is to establish a reliable baseline electromagnetic circuit dynamics so that any deviation observed in later sections can be attributed solely to plasma and secondary-electron-emission effects.
The circuit configuration used in the simulation is shown in Fig.~\ref{fig:Simplified Tesla transformer circuit}, and the corresponding parameters are listed in Table~\ref{tab:circuit_parameters}. 
The parameters were calibrated so that the model reproduces the experimentally measured output-capacitor voltage waveform prior to ion injection, thereby establishing a reference operating condition for subsequent plasma-coupled simulations. In Table~\ref{tab:circuit_parameters}, $k$ denotes the magnetic coupling coefficient between inductors $L_1$ and $L_2$. 
The initial condition assumes that capacitor $C_1$ is pre-charged to $3~\text{kV}$.
\begin{table}[t]
\centering
\caption{Circuit parameters of the simplified Tesla transformer model}
\label{tab:circuit_parameters}
\footnotesize
\begin{tabular}{|c c c|c c c|c c c|}
\hline
\multicolumn{3}{|c|}{\textbf{Resistors}} &
\multicolumn{3}{c|}{\textbf{Capacitors}} &
\multicolumn{3}{c|}{\textbf{Inductors \& Coupling}} \\
\hline
Param. & Value & Unit &
Param. & Value & Unit &
Param. & Value & Unit \\
\hline
$R_1$ & 5 & $\Omega$ &
$C_1$ & $3.8$ & $\mu$F &
$L_1$ & $47$ & $\mu$H \\

$R_2$ & 5 & $\Omega$ &
$C_2$ & $2.8$ & pF &
$L_2$ & 0.13 & H \\

$R_3$ & $100$ & $\text{M}\Omega$ &
$C_3$ & $0.2$ & pF &
$k$ & 0.37 & -- \\
\hline
\end{tabular}
\end{table}

\begin{figure}[t]
    \centering
    \includegraphics[width=0.75\linewidth]{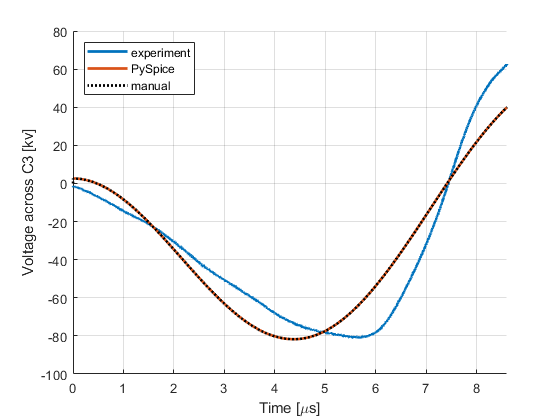}
    \caption{Comparison of the output capacitor (C3) voltage waveforms without ion source injection }
    \label{fig:C3_voltage_wo_ion}
\end{figure}
The charge-based manual circuit solver employing the BDF2 time-integration scheme  was validated against the commercial circuit simulator \texttt{PySpice}. 
Both simulations used an identical time step of $\Delta t = 15~\text{ns}$.
Because the experiment injects ions into capacitor C3, the voltage across C3 was chosen as the principal comparison quantity. 
Fig.~\ref{fig:C3_voltage_wo_ion} shows the simulated waveforms obtained from both solvers.
A quantitative comparison yields a root-mean-square error (RMSE) of only $0.006~\text{kV}$ between the two models, confirming that the proposed BDF2-based manual circuit update reproduces the dynamics of a commercial-grade circuit simulator with essentially indistinguishable accuracy. 
This agreement verifies both the charge formulation and the stability of the implicit time-marching scheme.

Next, the simplified circuit model was compared with experimental measurements, also shown in Fig.~\ref{fig:C3_voltage_wo_ion}. 
The simulated waveform exhibits a nearly sinusoidal response because the model consists exclusively of ideal lumped elements. 
In contrast, the measured waveform shows distortion caused by non-ideal high-voltage effects, including nonlinear switching behavior, parasitic inductance and capacitance in wiring, and dissipative losses in dielectric materials.
Nevertheless, the simulation accurately predicts the minimum voltage level and the time required to reach this minimum. 
Hence, the resonant frequency and global temporal evolution of the experimental signal are correctly captured.
These results demonstrate that the circuit model faithfully reproduces the intrinsic resonant dynamics of the Tesla transformer. 
Therefore, in the following sections, any additional waveform distortion or damping can be attributed to plasma formation and ion-induced SEE rather than circuit-model inaccuracies.

}

{\color{black}
\subsection{Tesla transformer with ion injection}

\begin{table}[t]
\centering
\footnotesize
\caption{Integrated simulation parameters}
\label{tab:space_time_variable}
\begin{tabular}{|c|c|c|}
\hline
\textbf{Param.} & \textbf{Value} & \textbf{Unit} \\
\hline
Simulation domain size (C3 gap size) &  $80$ & mm \\
Electrode area &  $2.53 \times 10^{-2}$ & m\textsuperscript{2} \\
Number of nodes & 601 & -- \\
Time segment $\Delta t$ & $15$ & ns \\
Total simulation time & $8.62$ & $\mu$s \\
Ion charge & $1.6 \times 10^{-19}$ & C \\
Electron charge & $-1.6\times 10^{-19}$ & C \\
Ion mass & $7.95 \times 10^{-26}$ & kg \\
(virtual) electron mass & $9.1 \times 10^{-28}$ & kg \\
Superparticle ratio & $2.5 \times 10^7$ & -- \\
\hline
\end{tabular}
\end{table}

The parameters used in the ion-injection simulation are summarized in Table~\ref{tab:space_time_variable}. 
Because the physical mass ratio between the \(\mathrm{Ti^+}\) ion and the electron is approximately \(8.7\times10^4\), direct use of the physical value leads to severe numerical stiffness in the particle simulation, as electron motion evolves orders of magnitude faster than ion transport. 
To alleviate this issue, a reduced mass ratio commonly adopted in PIC simulations was employed by increasing the electron mass by a factor of 1,000. 
This modification relaxes the electron plasma frequency while preserving the ion dynamics relevant to the discharge process.
Note that in a fully physical description without electron-mass scaling, energetic secondary electrons may reach tens of keV, for which their velocity becomes a non-negligible fraction of the speed of light. 
A more detailed kinetic treatment could therefore incorporate a relativistic particle pusher within an electrostatic PIC framework and, in principle, an electromagnetic PIC formulation.
However, the characteristic electron transit time in the gap is several orders of magnitude shorter than the circuit timescale governing the voltage evolution. 
Consequently, the circuit response is primarily determined by the time-averaged convection current rather than by sub-nanosecond electron oscillations. 
The present model can therefore be interpreted as a coarse-grained approximation in which fast electron dynamics are effectively averaged while preserving their net charge and current contribution to the circuit.
Moreover, although secondary electrons move much faster than ions, the net discharge current is primarily limited by the ion flux through the sheath; secondary electrons mainly provide current multiplication via SEE rather than independently determining the current magnitude.
\begin{figure}[t]
    \centering
    \includegraphics[width=1\linewidth]{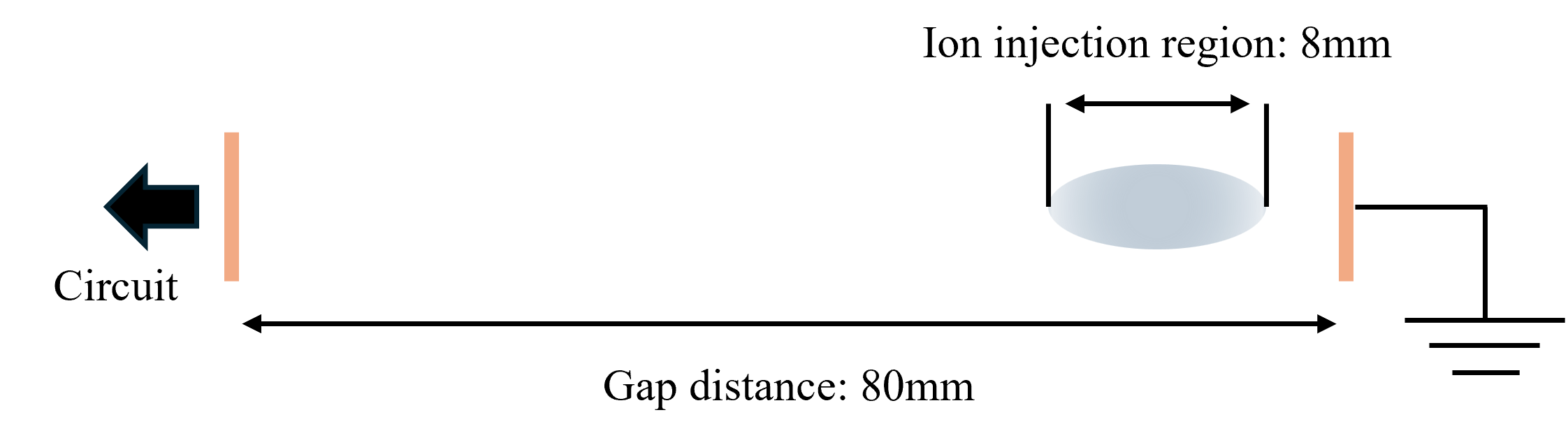}
    \caption{Ion injection region}
    \label{fig:ion_injection_region}
\end{figure}
\begin{figure}[t]
    \centering
    \includegraphics[width=0.8\linewidth]{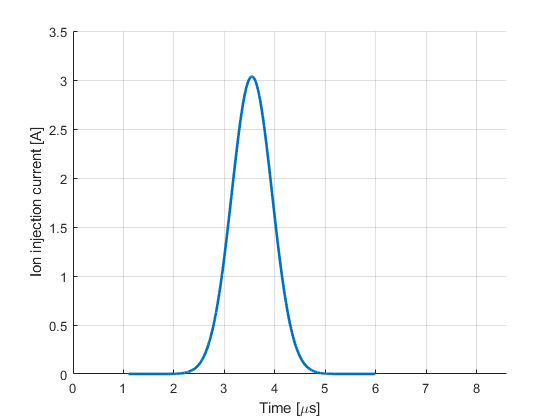}
    \caption{Ion injection current profile}
    \label{fig:ion_injection_current}
\end{figure}

\begin{figure}[t]
  \centering
  \includegraphics[width=0.75\linewidth]{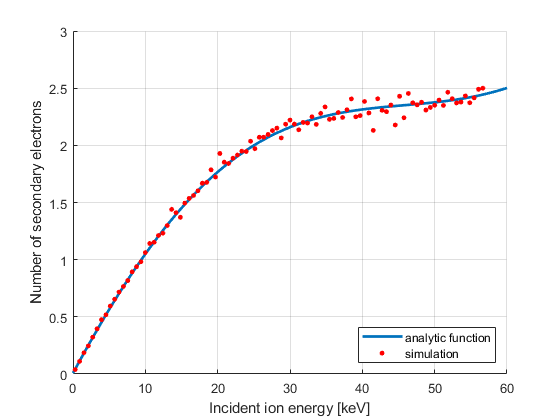}
  \caption{SEY as a function of incident ion energy, comparing Monte Carlo simulation results with an experiment-based fitting curve.}
  \label{fig:SE_yield_simulation}
\end{figure}

In the experiment, a plasma is generated inside a small cavity containing an aperture located near the grounded plate of the capacitor C3, and ions are injected into the gap through this aperture. 
To represent this configuration within a one-dimensional PIC framework, ion injection was modeled by introducing ions in a localized region adjacent to the grounded electrode, as depicted in Fig.~\ref{fig:ion_injection_region}. 
This treatment effectively averages the angular spread and geometric details of the aperture while preserving the net injected charge and current entering the discharge gap.
The temporal emission profile cannot be directly measured; therefore, the ion generation rate in the simulation was determined indirectly by matching the experimentally observed discharge behavior. 
Accordingly, the injected particle flux was prescribed as a Gaussian pulse, as illustrated in Fig.~\ref{fig:ion_injection_current}. Furthermore, as shown in Fig.~\ref{fig:SE_yield_simulation}, the SEY as a function of the incident ion energy is confirmed to closely follow the analytic function derived from the experimental data.

\begin{figure}[t]
    \centering
    \includegraphics[width=0.75\linewidth]{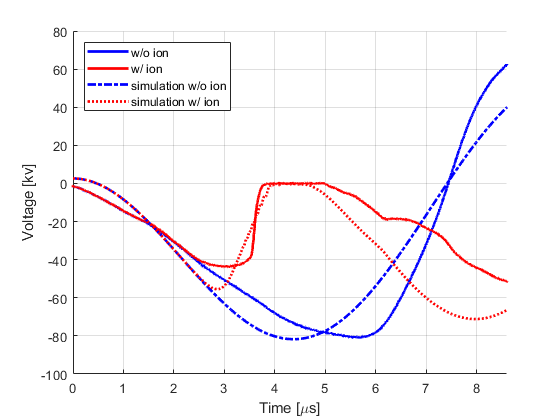}
    \caption{Comparison of the output capacitor voltage waveforms with ion source injection.} 
    \label{fig:C3_voltage_w_ion}
\end{figure}

\begin{figure}[t]
    \centering
    \includegraphics[width=0.95\linewidth]{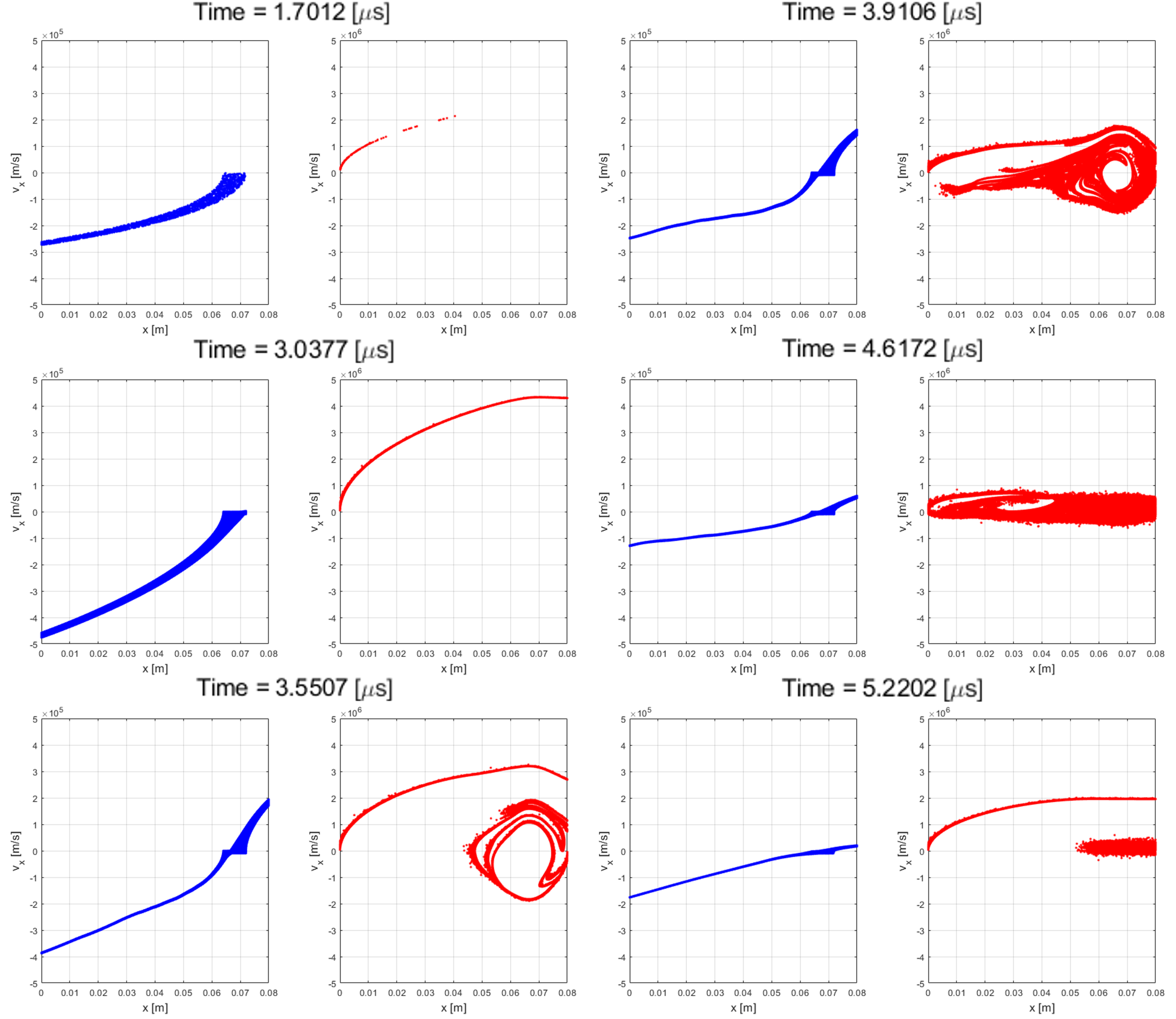}
    \caption{Temporal snapshots of particle dynamics in phase space at selected times
(blue: ions, red: secondary electrons).
For each time instant, ion distributions are shown in the left column and electron distributions in the right column.} 
    \label{fig:phase_domain}
\end{figure}

As shown in Fig.~\ref{fig:C3_voltage_w_ion}, the coupled simulation reproduces the experimentally observed abrupt voltage collapse followed by a sustained near-zero-voltage plateau. 
Such behavior cannot be obtained from the circuit model alone (Section~\ref{sec:result_1_w/o_ion}), indicating that the waveform distortion originates from plasma-induced current rather than intrinsic circuit nonlinearity.
To clarify the physical mechanism responsible for this evolution, Fig.~\ref{fig:phase_domain} presents time-resolved phase-space snapshots of ions and secondary electrons. 
These results show that the voltage waveform is governed by a feedback interaction between the plasma convection current and the external circuit response.
The discharge proceeds through five characteristic stages:
\begin{enumerate}
    \item At the onset of the discharge, injected ions accelerate toward the circuit-side electrode. 
    The resulting positive charge influx partially compensates the circuit current and delays the transition toward a more negative voltage.

    \item As the ion injection rate increases (Fig.~\ref{fig:ion_injection_current}), ions gain higher kinetic energy (Fig.~\ref{fig:incident_ion_energy}). 
    The SEY exceeds two (Fig.~\ref{fig:SE_yield_simulation}), initiating strong electron multiplication at the electrode surface.

    \item The rapid growth of convection current overwhelms the circuit current, causing an abrupt collapse of the capacitor voltage. 
    As the voltage drops, ion--ion repulsion redistributes particle energies and induces SEE at both electrodes, leading to transient plasma instabilities.

    \item When the capacitor voltage approaches zero, ions impacting both electrodes acquire nearly identical energies (Fig.~\ref{fig:phase_domain}). 
    Despite the vanishing electric potential difference, a finite convection current persists, thereby maintaining the near-zero-voltage plateau.

    \item As the ion injection rate decreases, both ion density and impact energy diminish, reducing SEE. 
    The circuit current then regains dominance, and the system gradually returns to its natural resonant trajectory.
\end{enumerate}

\begin{figure}[t]
  \centering
  \includegraphics[width=0.75\linewidth]{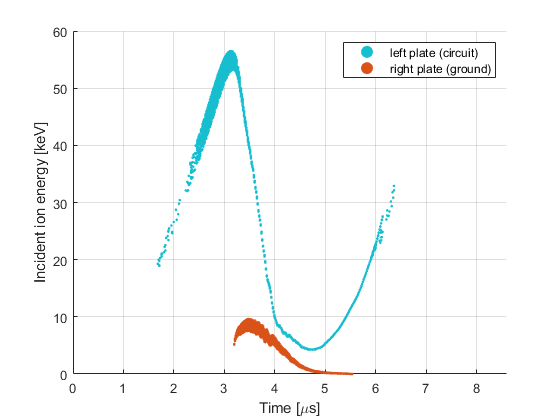}
  \caption{Incident ion energy over time.}
  \label{fig:incident_ion_energy}
\end{figure}

These observations indicate that the plateau voltage is not a passive post-breakdown state but a dynamically sustained equilibrium. 
The plasma-generated convection current counterbalances the circuit current, establishing a self-regulated circuit--plasma operating point. 
As the ion injection weakens, this balance is lost and the system gradually returns to its intrinsic circuit resonance.

}

{\color{black}
\subsection{Comparison of weak and strict coupling modes}
\begin{figure}[t]
    \centering
    \includegraphics[width=0.75\linewidth]{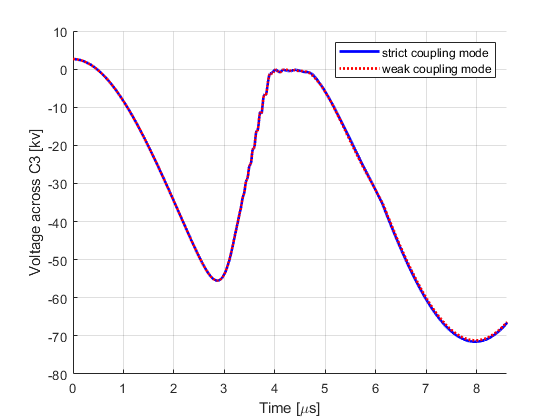}
    \caption{Comparison of the output capacitor voltage waveforms without ion source injection (the weak-coupling result is shown as a dashed line).}
    \label{fig:strict_weak}
\end{figure}
To assess the system-level impact of the coupling strategy, simulation results obtained using the strict and weak coupling modes are compared in Fig.~\ref{fig:strict_weak}.
The two voltage waveforms exhibit close agreement throughout the entire simulation interval, with an RMSE of \SI{0.228}{\kilo\volt}. 
This indicates that the weak coupling mode accurately captures the global circuit dynamics despite its reduced coupling complexity. 
In other words, resolving the circuit--plasma interaction at every sub-time-step is not necessary to reproduce the macroscopic voltage evolution.
These results suggest that, for a broad class of practical circuits, the \texttt{PySpice}-based weak coupling mode can reproduce the macroscopic voltage response with sufficient accuracy, enabling efficient simulation without fully resolving the tight circuit–plasma coupling at every sub-time-step.

A detailed comparison between the weak and strict coupling modes at the local time-step scale is provided in \ref{App_comp_strict_weak}.

}

{\color{black}
\subsection{Limitation of SEE-free simulation}

In the previous section, it was demonstrated that the simulation incorporating the proposed SEE module successfully reproduces the expected physical behavior. Accordingly, this section examines the system response under simulation conditions in which SEE modeling is removed. To this end, simulations were performed with SEE selectively disabled while all other parameters were kept identical to those used in the previous section (Tables~\ref{tab:circuit_parameters} and~\ref{tab:space_time_variable}), except for electron-related parameters, which were excluded since electrons do not appear in the present simulations. 

\begin{figure}
  \centering
  \includegraphics[width=0.8\linewidth]{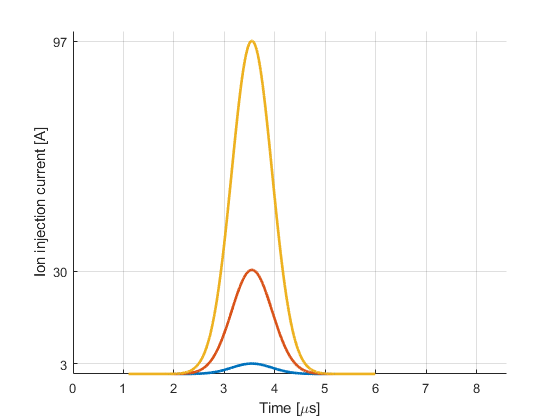}
  \caption{Ion injection current profiles for three input-current cases used in the SEE-free simulations.}
  \label{fig:ion_injection_current_2}
\end{figure}

\begin{figure}
  \centering
  \includegraphics[width=0.8\linewidth]{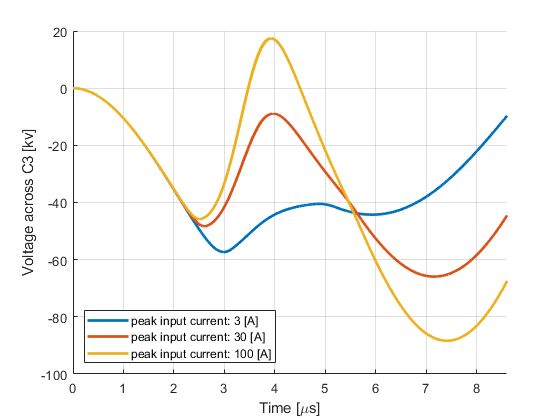}
  \caption{Ion injection current profiles for three input-current cases used in the SEE-free simulations.}
  \label{fig:C3_voltage_wo_SEE}
\end{figure}

\begin{figure}
  \centering
  \includegraphics[width=0.8\linewidth]{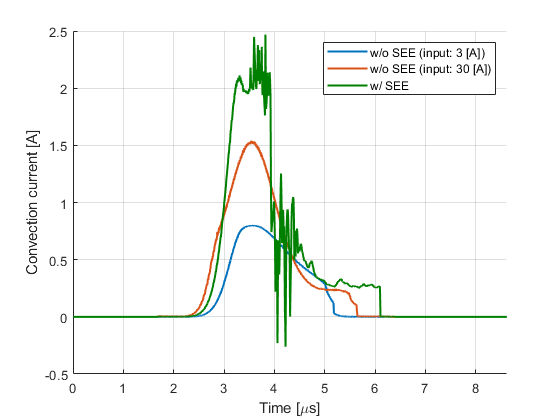}
  \caption{Time evolution of the convection current comparing a simulation with SEE at a peak input current of \SI{3}{\ampere} and without SEE at peak input currents of 3 and \SI{30}{\ampere}.}
  \label{fig:convection_current}
\end{figure}

As an initial test, the same input current employed in the SEE-included case was applied, corresponding to the blue curve in Fig.~\ref{fig:ion_injection_current_2}. The resulting capacitor voltage evolution, shown in Fig.~\ref{fig:C3_voltage_wo_SEE}, exhibits no voltage breakdown, indicating that ion injection alone is insufficient to drive the electrode potential down to \SI{0}{\volt}. 

To examine whether a voltage breakdown can be observed in the ion-only simulations by increasing the ion injection rate, an additional simulation was performed using the input current corresponding to the red curve in Fig.~\ref{fig:ion_injection_current_2}. In this case, the effective ion injection current was increased by raising the superparticle weight while keeping the total number of particles fixed. Nevertheless, despite the increased ion-injection, the capacitor voltage still does not collapse to \SI{0}{\volt}, as indicated by the red trace in Fig.~\ref{fig:C3_voltage_wo_SEE}, even under this enhanced injection condition. 

This absence of voltage collapse can be attributed to the reduced convection current observed in the ion-only case. As shown in Fig.~\ref{fig:convection_current}, the convection current in the time window of approximately \SIrange{3}{4}{\micro\second} remains significantly lower than that obtained when SEE is included, indicating insufficient charge transfer to trigger a complete voltage breakdown.

Finally, an extreme ion injection condition was considered, corresponding to the yellow curve in Fig.~\ref{fig:ion_injection_current_2}. In this case, a pronounced voltage breakdown is observed (Fig.~\ref{fig:C3_voltage_wo_SEE}), during which the capacitor voltage rapidly transitions from approximately $-$\SI{50}{\kilo\volt} to $+$\SI{20}{\kilo\volt}. However, in contrast to the experimental observation shown in Fig.~\ref{fig:C3_voltage_w_ion}, the simulated voltage does not exhibit a sustained near-zero-voltage plateau. This result demonstrates that, while sufficiently intense ion injection can trigger a voltage breakdown, the maintenance of the post-breakdown voltage state critically depends on the stabilizing contribution of SEE.

}
}

\section{Conclusion}
In this work, a self-consistent circuit--plasma co-simulation framework
has been developed and applied to investigate voltage breakdown
phenomena in high-voltage pulsed vacuum systems.
The proposed approach explicitly couples a lumped-element external circuit,
represented by a Tesla transformer configuration, with a one-dimensional
electrostatic PIC solver describing charged-particle dynamics inside
a vacuum capacitor gap.
By resolving the bidirectional interaction between circuit-level charge
transport and plasma-induced convection currents, the framework provides
a unified time-marching description of macroscopic circuit dynamics
and microscopic plasma processes.

A central contribution of this study is the explicit incorporation
of ion-induced secondary electron emission (SEE) into the
circuit--plasma coupling.
The SEE process is modeled using a Monte Carlo procedure
based on experimentally informed yield and energy distributions,
allowing secondary electrons to be generated self-consistently
in response to energetic ion impacts on electrode surfaces.
The emitted electrons substantially enhance the convection current
at the electrode, modify the surface charge density,
and directly feed back into the external circuit response.
This SEE-aware formulation extends conventional circuit--plasma
co-simulation approaches that consider only primary particle transport.

The numerical results demonstrate that SEE plays a decisive role
in the voltage breakdown dynamics.
When SEE is included, the model reproduces the experimentally observed
rapid voltage collapse followed by a sustained near-zero-voltage plateau.
Analysis of the coupled dynamics shows that high-energy ion impact
during peak injection produces a strong increase in SEE yield,
which in turn sharply amplifies the convection current and drives
the abrupt voltage breakdown.
In contrast, simulations performed without SEE fail to produce
voltage collapse under comparable ion injection conditions,
even when the ion flux is artificially increased.
Although extreme ion injection can induce a transient voltage drop
in SEE-free simulations, the absence of sustained post-breakdown
behavior confirms the essential role of SEE in maintaining
the near-zero-voltage state.

To enhance interoperability with practical circuit-analysis tools,
a weak coupling strategy was introduced in addition to the fully
implicit strict coupling formulation.
In the weak coupling mode, the external circuit is solved
using a SPICE-based solver (here \texttt{PySpice}),
and the resulting circuit-side charge is sequentially coupled
to the plasma module.
Despite the partitioned treatment, the weak coupling approach
reproduces the same system-level voltage breakdown characteristics
as the strict coupling formulation, including breakdown onset,
voltage collapse, and post-breakdown evolution.
Quantitative comparisons show only minor discrepancies
at the local time-step level.
This result indicates that accurate circuit--plasma--SEE interaction
can be captured without direct access to the internal circuit equations,
provided that the global time step adequately resolves
the charge-variation timescale.

Overall, the present study establishes that predictive simulation
of high-voltage breakdown in vacuum systems requires a fully coupled
treatment of circuit dynamics, plasma evolution,
and surface emission processes.
The proposed SEE-enabled circuit--plasma framework offers
a flexible and extensible platform for integrating detailed
plasma and surface physics with realistic external circuit models,
and provides a physically grounded basis for the analysis
and design of advanced high-power pulsed vacuum systems.

%% The Appendices part is started with the command \appendix;
%% appendix sections are then done as normal sections

\appendix
{\color{black}
\section{Detailed formulation of the manual circuit solver}
\label{App_circuit}

This appendix presents the complete algebraic formulation of the manual
KVL/KCL-based circuit solver introduced in Section~\ref{sec:manual_solver}.
The continuous-time loop equations are discretized using the BDF2 scheme,
and the resulting linear system is derived explicitly.

\subsection{BDF2 discretization of the loop equations}

Starting from the loop equations
Eqs.~\eqref{eq:loop_1}--\eqref{eq:loop_5}, 
the first- and second-order time derivatives are approximated using
the second-order backward differentiation formulas

\begin{align}
\frac{dQ}{dt}\biggr|^n
&\approx
\frac{3Q^n-4Q^{n-1}+Q^{n-2}}{2\Delta t},
\\
\frac{d^2Q}{dt^2}\biggr|^n
&\approx
\frac{9Q^n - 24Q^{n-1} + 22Q^{n-2}
- 8Q^{n-3} + Q^{n-4}}
{4\Delta t^2}.
\end{align}
Substituting these expressions into the loop equations and collecting
terms evaluated at time level $n$ yields a linear algebraic system
\begin{equation}
\overline{\mathbf{T}} \cdot \mathbf{Q}^n
=
\mathbf{C},
\label{eq:matrix_form_app}
\end{equation}
where $\mathbf{Q}^n =[Q_1^n, Q_2^n, Q_3^n, Q_4^n, Q_5^n]^T$.

\subsection{Coefficient matrix}
The coefficient matrix has the block structure

\begin{equation}
\overline{\mathbf{T}} =
\begin{bmatrix}
\overline{\mathbf{T}}_{1} & \overline{\mathbf{T}}_{2} \\
\overline{\mathbf{T}}_{3} & \overline{\mathbf{T}}_{4}
\end{bmatrix}.
\end{equation}
The leading $3\times3$ block associated with
$(Q_1,Q_2,Q_3)$ is
\begin{align}
\overline{\mathbf{T}}_{1} =
\begin{bmatrix}
\frac{3R_1}{2\Delta t}+\frac{1}{C_1} &
-\frac{3R_1}{2\Delta t} & 0
\\
-\frac{3R_1}{2\Delta t} &
\frac{3R_1}{2\Delta t}
+\frac{3R_2}{2\Delta t}
+\frac{9L_1}{4\Delta t^2}
&
\frac{9M}{4\Delta t^2}
\\
0 &
\frac{9M}{4\Delta t^2} &
\frac{9L_2}{4\Delta t^2}
+\frac{3R_3}{2\Delta t}
\end{bmatrix}.
\end{align}
The remaining block associated with $(Q_4,Q_5)$ is
\begin{align}
\overline{\mathbf{T}}_{4} =
\begin{bmatrix}
\frac{3R_3}{2\Delta t}+\frac{1}{C_3}
&
-\frac{3R_3}{2\Delta t}
\\
-\frac{3R_3}{2\Delta t}
&
\frac{3R_1}{2\Delta t}
+\frac{3R_3}{2\Delta t}
+\frac{1}{C_2}
\end{bmatrix}.
\end{align}
The off-diagonal coupling blocks
$\overline{\mathbf{T}}_{2}$ and $\overline{\mathbf{T}}_{3}$
originate from resistor interactions between loops:
\begin{align}
\overline{\mathbf{T}}_{2}
=
\overline{\mathbf{T}}_{3}^T
=
\begin{bmatrix}
0 & 0
\\
-\frac{3R_1}{2\Delta t} &
\frac{3R_1}{2\Delta t}
\\
0 &
\frac{3R_3}{2\Delta t}
\end{bmatrix}.
\end{align}

\subsection{Right-hand-side vector}
The vector $\mathbf{C} = [c_1, c_2, c_3, c_4, c_5]^T$ contains all known contributions from previous
time steps arising from the BDF2 discretization.
Each component $c_i$ is obtained by substituting the BDF2 expressions
into the loop equations and isolating terms that depend solely on
previous time steps $(n-1,n-2,n-3,n-4)$ given by
\begin{align}
c_1 &= \frac{R_1}{2\Delta t} \Big(
-4Q_1^{n-1} + Q_1^{n-2}
+ 4Q_2^{n-1} - Q_2^{n-2} 
+ 4Q_5^{n-1} - Q_5^{n-2}
\Big),\\
c_2 &= \frac{R_1}{2\Delta t} \Big(
-4Q_1^{n-1} + Q_1^{n-2}
+ 4Q_2^{n-1} - Q_2^{n-2} 
+ 4Q_5^{n-1} - Q_5^{n-2}
\Big) \notag \\
&+ \frac{R_2}{2\Delta t} \Big(
4Q_2^{n-1} - Q_2^{n-2}
\Big) \notag \\
&+ \frac{L_1}{4\Delta t^2} \Big(
24Q_2^{n-1} + 22Q_2^{n-2}
+ 8Q_2^{n-3} + Q_2^{n-4}
\Big)\notag \\
&+ \frac{M}{4\Delta t^2} \Big(
24Q_3^{n-1} + 22Q_3^{n-2}
+ 8Q_3^{n-3} + Q_3^{n-4}
\Big),\\
c_3 &= \frac{R_3}{2\Delta t} \Big(
4Q_3^{n-1} - Q_3^{n-2}
+ 4Q_4^{n-1} - Q_4^{n-2} + 4Q_5^{n-1} - Q_5^{n-2}
\Big) \notag \\
&+ \frac{L_2}{4\Delta t^2} \Big(
24Q_3^{n-1} + 22Q_3^{n-2}
+ 8Q_3^{n-3} + Q_3^{n-4}
\Big) \notag \\
&+ \frac{M}{4\Delta t^2} \Big(
24Q_2^{n-1} + 22Q_2^{n-2}
+ 8Q_2^{n-3} + Q_2^{n-4}
\Big),\\
c_4 &= \frac{R_3}{2\Delta t} \Big(
4Q_3^{n-1} - Q_3^{n-2}
+ 4Q_4^{n-1} - Q_4^{n-2} + 4Q_5^{n-1} - Q_5^{n-2}
\Big),\\
c_5 &= \frac{R_3}{2\Delta t} \Big(
4Q_3^{n-1} - Q_3^{n-2}
+ 4Q_4^{n-1} - Q_4^{n-2} + 4Q_5^{n-1} - Q_5^{n-2}
\Big) \notag \\
& + \frac{R_1}{2\Delta t} \Big(
-4Q_1^{n-1} - Q_1^{n-2}
+ 4Q_2^{n-1} - Q_2^{n-2} + 4Q_5^{n-1} - Q_5^{n-2}
\Big).
\end{align}

\section{Detailed formulation of the ES-PIC plasma solver}
\label{App_plasma}

This appendix provides the full discretization details of the one-dimensional ES-PIC solver used in the multiscale circuit--plasma framework.

The computational domain represents the one-dimensional region between 
the two electrodes of the load capacitor $\mathrm{C3}$.
\begin{figure}[t]
    \centering
    \includegraphics[width=0.9\linewidth]{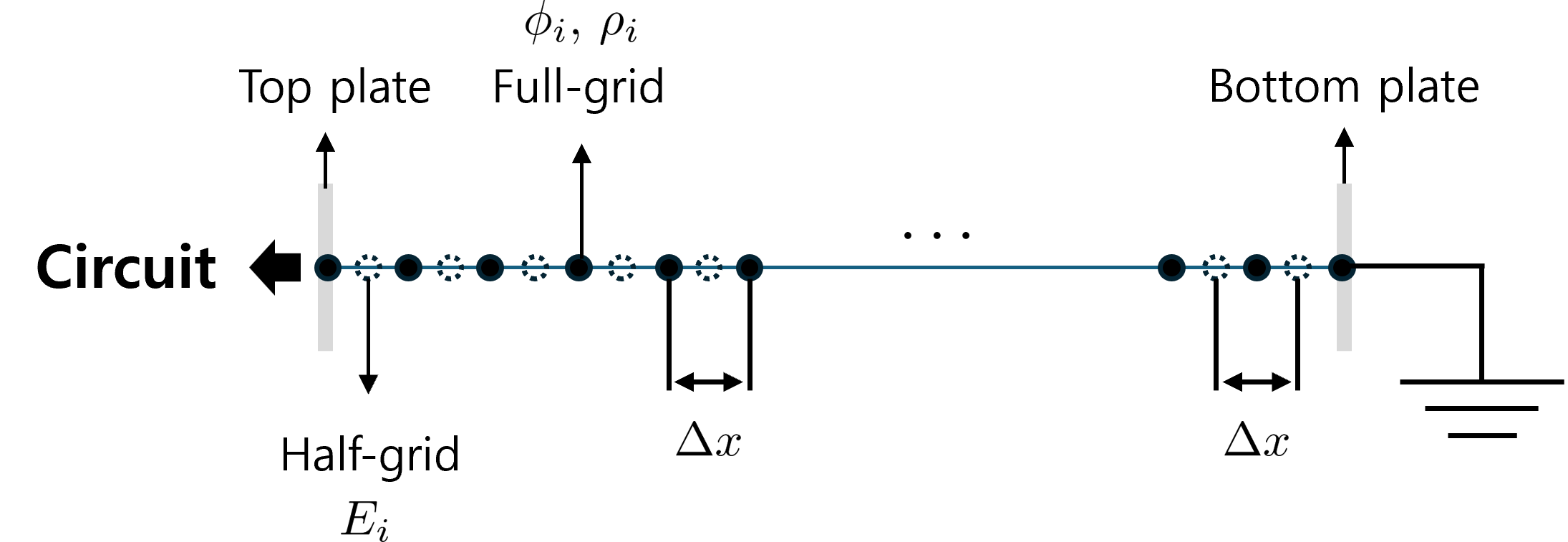}
    \caption{Schematic of the one-dimensional kinetic plasma solver domain.
    The electric potential $\phi$ and charge density $\rho$ are defined at nodes,
    while the electric field $E$ is defined at half-grid locations.}
    \label{fig:1D_domain_app}
\end{figure}
The domain is discretized using a uniform grid with spacing $\Delta x$.
The electric potential $\phi$ and charge density $\rho$ are defined at grid nodes,
whereas the electric field is defined at half-grid locations between adjacent nodes,
as illustrated in Fig.~\ref{fig:1D_domain_app}.
From the perspective of differential geometry, $\phi$ is a 0-form,
$\mathbf{E}$ is a 1-form, and $\rho$ is a 3-form.
In the discrete setting, the Hodge dual of $\rho$ is treated as a 0-form.
Therefore, $\phi$ and $\rho$ are assigned to nodes, while $\mathbf{E}$ is assigned to edges.

The governing equation is the Poisson equation
\begin{equation}
\nabla^2 \phi = -\frac{\rho}{\epsilon_0}.
\end{equation}
Using the central difference scheme, the discretized form becomes
\begin{equation}
\frac{\phi_{i+1}^n - 2\phi_i^n + \phi_{i-1}^n}{\Delta x^2}
=
-\frac{\rho_i^n}{\epsilon_0}.
\label{eq:dt_Poisson_app}
\end{equation}
The electric field at half-grid locations is computed as
\begin{equation}
E_{i+\frac{1}{2}}^n
=
\frac{\phi_i^n - \phi_{i+1}^n}{\Delta x}.
\label{eq:dt_E-field}
\end{equation}
The lower electrode is grounded:
\begin{equation}
\phi_N^n = 0.
\end{equation}

At the upper electrode, Gauss’s law yields
\begin{equation}
\epsilon_0 E_{\frac{1}{2}}^n
=
\sigma_+^n
+
\frac{\Delta x}{2}\rho_0^n,
\label{eq:1D_Gaussian_app}
\end{equation}
where $\sigma_+^n$ denotes the surface charge supplied by the external circuit.
Substituting Eq.~\eqref{eq:dt_E-field}, we obtain
\begin{equation}
-\phi_0^n + \phi_1^n
=
-\frac{\Delta x^2}{\epsilon_0}
\left(
\frac{\sigma_+^n}{\Delta x}
+
\frac{\rho_0^n}{2}
\right).
\end{equation}

The discrete Poisson equation can be written in matrix form:
\begin{equation}
\overline{\mathbf{L}} \cdot \boldsymbol{\phi}^n
=
\boldsymbol{\rho}^n,
\end{equation}
where the discrete Laplacian operator is

\[
\overline{\mathbf{L}} =
\begin{bmatrix}
-1 & 1 & 0 & \cdots & 0 \\
1 & -2 & 1 & \cdots & 0 \\
0 & 1 & -2 & \cdots & 0 \\
\vdots & \vdots & \vdots & \ddots & \vdots \\
0 & 0 & 0 & \cdots & -2
\end{bmatrix}.
\]
The first row incorporates the modified boundary condition at the upper electrode.

\begin{figure}[t]
    \centering
    \includegraphics[width=0.6\linewidth]{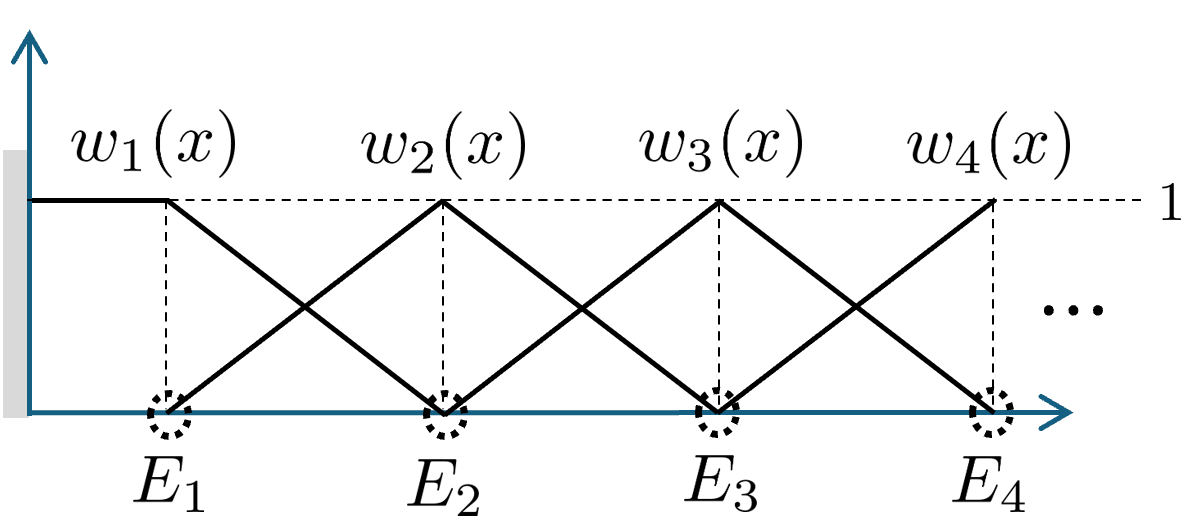}
    \caption{One-dimensional linear weighting (cloud-in-cell) scheme.
    The particle field value is interpolated from adjacent half-grid points.}
    \label{fig:E_CIC_app}
\end{figure}

\section{Field interpolation and charge deposition}
\label{App_interpolation_deposition}
Since particles are not restricted to grid points, the electric field at
particle positions is obtained via linear interpolation (cloud-in-cell method):
\begin{equation}
E_p^n
=
w_i(x_p^n) E_{i+\frac{1}{2}}^n
+
w_{i+1}(x_p^n) E_{i+\frac{3}{2}}^n,
\end{equation}
where $w_i(x)$ denotes the linear weighting function illustrated in
Fig.~\ref{fig:E_CIC_app}.
Charge deposition to the grid is performed using the same weighting scheme,
ensuring discrete charge conservation.
This completes the detailed formulation of the ES-PIC plasma solver used in
the multiscale circuit--plasma simulation framework.

\section{Time segment selection and stability considerations}
\label{App_time_segment}
The proposed SEE-enabled circuit--plasma co-simulation framework
involves multiple subsystems characterized by distinct temporal scales:
(i) resonant LC dynamics of the external circuit,
(ii) electrostatic plasma response inside the load capacitor,
and (iii) kinetic motion of charged particles.
The global time step $\Delta t$ must therefore satisfy both
stability and resolution requirements across all modules.

\subsection{Circuit-resolution constraint}
The circuit equations are discretized using the second-order
backward differentiation formula (BDF2),
which is unconditionally stable for linear stiff systems.
Thus, stability does not directly restrict $\Delta t$.
Instead, $\Delta t$ must resolve the highest circuit resonance
frequency $\omega_{\max}$:
\begin{equation}
\Delta t \le \frac{1}{N_{\omega}\,\omega_{\max}},
\end{equation}
where $N_{\omega}$ denotes the number of time steps per oscillation period.
In this study, $N_{\omega}=20$--$50$
to limit numerical phase errors in transient LC oscillations.

\subsection{PIC cell-crossing constraint}

Although the Poisson equation is solved implicitly,
particle trajectories are advanced explicitly via the leapfrog scheme.
To ensure accurate charge deposition and field interpolation,
particles must not traverse excessive grid distances within one time step:
\begin{equation}
\Delta t \le \beta \frac{\Delta x}{v_{\max}},
\label{eq:cell_crossing_revised}
\end{equation}
where $v_{\max}$ is the maximum particle velocity
and $\beta=0.2$--$0.5$ is a safety factor.

\subsection{Electron plasma-frequency constraint}

When secondary electrons are explicitly advanced,
an additional constraint arises from the electron plasma frequency:
\begin{equation}
\Delta t \le \alpha \,\omega_{pe}^{-1},
\label{eq:plasma_frequency_revised}
\end{equation}
where $\alpha=0.1$--$0.2$.
Since the secondary-electron density emerges self-consistently,
this condition is evaluated \emph{a posteriori}
using the maximum simulated electron density.

\subsection{Coupling-accuracy constraint}

The circuit and plasma modules are implicitly coupled through
$\sigma_+^n$ and $\phi_0^n$.
To preserve quasi-static consistency of the Poisson solve,
the charge variation per time step must remain moderate:
\begin{equation}
\Delta t \le
\eta \frac{Q_{\mathrm{ref}}}{I_{\mathrm{ref}}},
\end{equation}
with $\eta=0.05$--$0.2$.

\subsection{Final time-step selection}

The global time step is chosen as
\begin{equation}
\Delta t
=
\min
\left(
\frac{1}{N_{\omega}\omega_{\max}},
\;
\beta \frac{\Delta x}{v_{\max}},
\;
\alpha \omega_{pe}^{-1},
\;
\eta \frac{Q_{\mathrm{ref}}}{I_{\mathrm{ref}}}
\right).
\end{equation}
In practice, the PIC constraints are more restrictive than
the circuit-resolution constraint.
Accordingly, $\Delta t$ is primarily determined by
Eqs.~\eqref{eq:cell_crossing_revised}
and~\eqref{eq:plasma_frequency_revised}.

\subsection{A posteriori verification}

Because the secondary-electron density and velocity are not known
\emph{a priori}, preliminary simulations were performed to estimate
$v_{\max}$ and $n_e$.
The selected time step,
$\Delta t = 1.5\times10^{-11}\,\mathrm{s}$,
satisfies both the plasma-frequency and cell-crossing conditions.

At the moment of maximum electron density
(Fig.~\ref{fig:max_chrge_num}),
the plasma frequency was
$\omega_{pe,\max}\approx 6.6\times10^9\,\mathrm{rad/s}$,
ensuring compliance with
Eq.~\eqref{eq:plasma_frequency_revised}.
The maximum electron velocity was
$v_{e,\max}\approx 4.5\times10^6\,\mathrm{m/s}$,
yielding a displacement
$v_{e,\max}\Delta t \approx 6.75\times10^{-5}\,\mathrm{m}$,
which remains below the grid spacing
$\Delta x=1.33\times10^{-4}\,\mathrm{m}$.

\subsection{Spatial resolution}
To maintain electrostatic accuracy,
the relative electric-field variation within one cell must satisfy
\begin{equation}
\frac{|\Delta E|}{|E|}
\approx
\frac{|\rho| \Delta x}{\epsilon_0 |E|}
\ll 1,
\end{equation}
ensuring that the discretized Poisson solution resolves local charge-induced field gradients.

\begin{figure}
    \centering
    \includegraphics[width=0.8\linewidth]{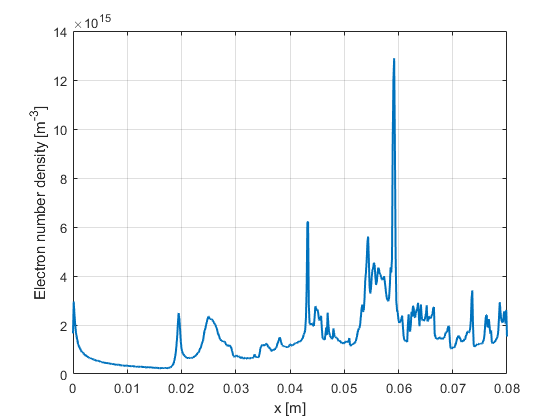}
    \caption{Snapshot of the electron distribution in the capacitor gap at the moment when the local electron density reaches its maximum in the simulation}
    \label{fig:max_chrge_num}
\end{figure}

}

\section{Validation}
\label{App_validation}

\begin{table}[htbp]
\centering
\footnotesize 
\caption{simulation parameters used for validation}
\label{tab:validation_parameters}
\begin{tabular}{|c|c|c|l|}
\hline
\textbf{Parameter} & \textbf{Value} & \textbf{Unit} \\
\hline
\(\mathrm{R_1}\) &  $1.8\times 10^{4}$ & [\SI{}{\ohm}] \\
\(\mathrm{C_1}\) &  $2.8 \times 10^{-12}$ & [\SI{}{\farad}] \\
Simulation domain size (Capacitor gap size) &  $8 \times 10^{-2}$ & [\SI{}{\meter}] \\
Electrode area &  $2.53 \times 10^{-2}$ & [\SI{}{\meter^2}] \\
Number of nodes & 601 & [-] \\
Time step & $1.5 \times 10^{-11}$ & [\SI{}{\second}] \\
Total simulation time & $8 \times 10^{-8}$ & [\SI{}{\second}] \\
ion charge & $1.6 \times 10^{-19}$ & [\SI{}{\coulomb/\meter^2}] \\
ion mass & $7.95 \times 10^{-26}$ & [\SI{}{\kilo\gram}] \\
superparticle ratio & $2.5 \times 10^7$ & [-] \\
\hline
\end{tabular}
\end{table}

\begin{figure}
    \centering
    \includegraphics[width=0.7\linewidth]{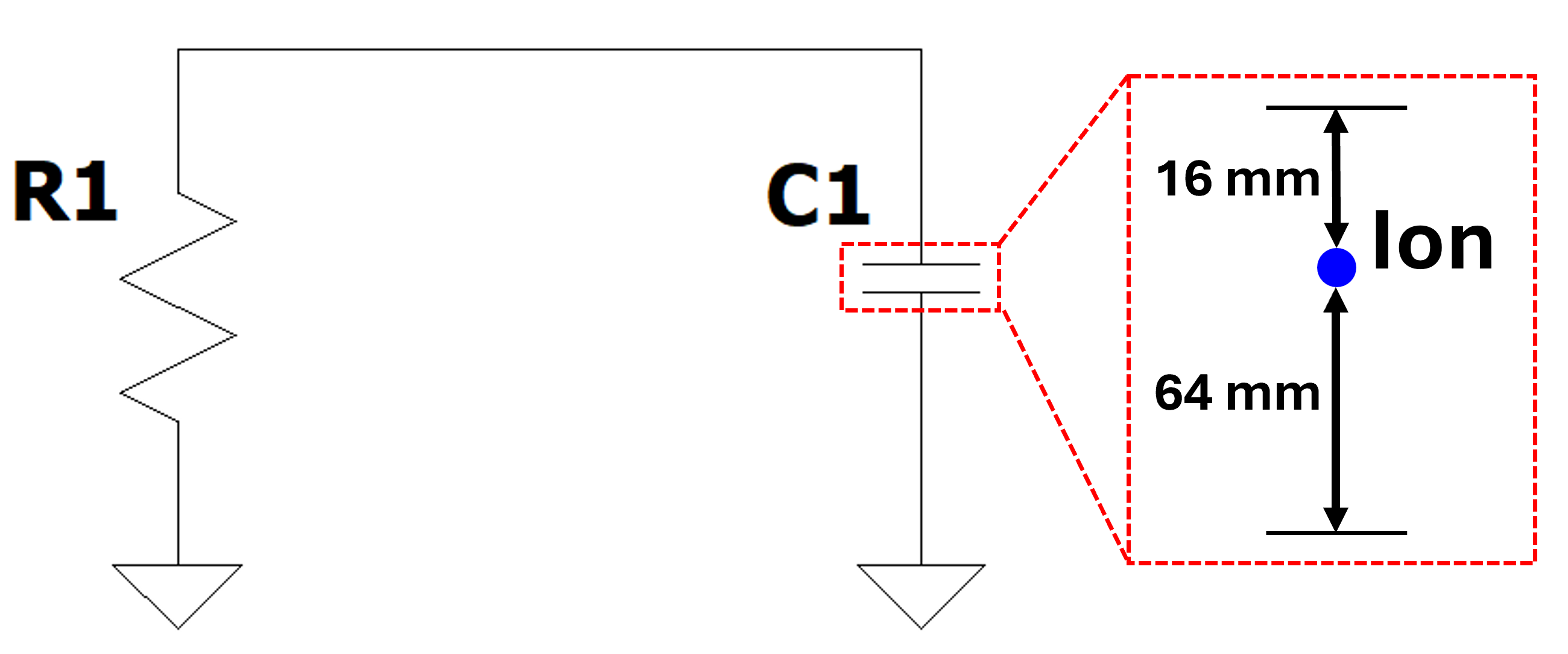}
    \caption{RC circuit model used for validation of the integrated simulation}
    \label{fig:validation_circuit}
\end{figure}

To evaluate the capability of the circuit-plasma integrated simulation to accurately model physical phenomena, a simulation was performed on a simple circuit for which physical behavior can be predicted. All simulations used for validation were conducted using a common set of parameters, as summarized in Table~\ref{tab:validation_parameters}.

\subsection{Stationary ion injection}

The first case considered involves generating a stationary ion in the gap of capacitor C1 in a simple RC circuit, as illustrated in Fig.~\ref{fig:validation_circuit}. Since no external voltage source is applied and the capacitor is assumed to be initially fully discharged, the initial potential distribution in the capacitor gap is zero at all spatial locations. At \(t=\SI{7.5}{\nano\second}\), a stationary ion is introduced at a specified position within the capacitor gap; then the electric potential \(\phi(x)\) in the one-dimensional determined by
\begin{equation}
\label{eq:1D_Poisson}
\frac{d^2 \phi}{dx^2} = -\frac{\rho(x)}{\epsilon_0},
\end{equation}
where
\begin{equation}
\label{eq:delta_source}
\rho(x) = q_\text{gap}\cdot\delta(x - x_0).
\end{equation}
Here, \(q_\text{gap}\) denotes the surface charge density of the equivalent charge sheet located within the capacitor gap.
Following Eqs.~\eqref{eq:1D_Poisson} and~\eqref{eq:delta_source} the solution of the Poisson equation can be obtained by considering the properties of the delta-function charge distribution. For all positions except at the ion location \(x\neq x_0\), the charge density vanishes, and Eq.~\eqref{eq:1D_Poisson} reduces to
\begin{equation}
\label{eq:1D_Poisson_2}
\frac{d^2 \phi}{dx^2} = 0,
\end{equation}
indicating that the electric potential is a linear function of \(x\) in each subdomain separated by the ion position. Accordingly, the potential can be expressed in a piecewise form as
\begin{equation}
\label{eq:potential_left_right}
\phi(x) =
\begin{cases}
A_1 x + B_1, & x < x_0, \\
A_2 x + B_2, & x > x_0,
\end{cases}
\qquad
\phi(x_0^-)=\phi(x_0^+).
\end{equation}
where the coefficients \(A_1\),  \(A_2\),  \(B_1\) and \(B_2\) are determined from boundary and continuity conditions.
The grounded electrode located at \(x=L\) imposes a Dirichlet boundary condition, 
\begin{equation}
\phi(L)=0.
\end{equation}
To close the system, a homogeneous Neumann boundary condition is applied at the opposite boundary, corresponding to a vanishing electric field: 
\begin{equation}
\left.\frac{d \phi}{d x}\right|_{x = 0} = 0.
\end{equation}
Under these conditions, the coefficients in Eq.~\eqref{eq:potential_left_right} can be uniquely determined, yielding the potential distribution
\begin{equation}
\label{eq:potential_left_right_2}
\phi(x) =
\begin{cases}
\dfrac{q_\text{gap}}{\epsilon_0}\left(L - x_0\right), & \text{for } x < x_0, \\[3ex]
\dfrac{q_\text{gap}}{\epsilon_0}\left(L - x\right),   & \text{for } x > x_0.
\end{cases}
\end{equation}
\begin{figure}
    \centering
    \includegraphics[width=0.75\linewidth]{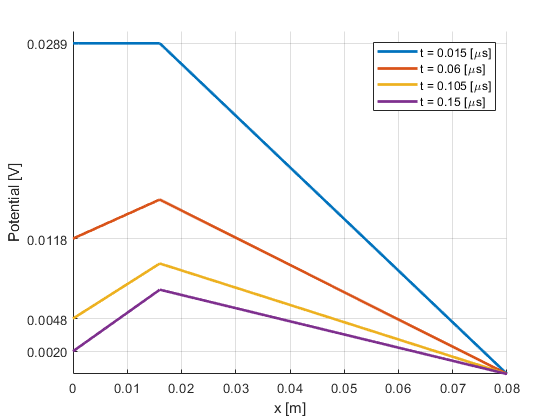}
    \caption{Potential distribution in the capacitor gap of the RC circuit with stationary ion between the plates.}
    \label{fig:validation_potential_1}
\end{figure}

The simulation results demonstrate that, immediately following the injection of an ion at the location shown in Fig.~\ref{fig:validation_circuit}, the electric potential assumes a distribution that satisfies Eq.~\eqref{eq:potential_left_right_2}, as shown by the blue line in Fig.~\ref{fig:validation_potential_1}. 
\begin{figure}
    \centering
    \includegraphics[width=0.75\linewidth]{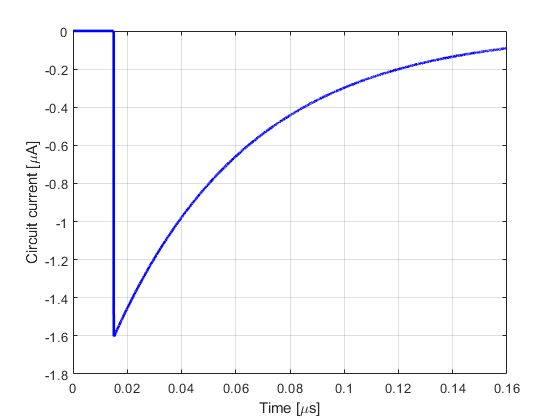}
    \caption{Time evolution of the circuit current in the RC circuit with stationary ion between the plates.}
    \label{fig:validation_current_1}
\end{figure}
\begin{figure}
    \centering
    \includegraphics[width=0.75\linewidth]{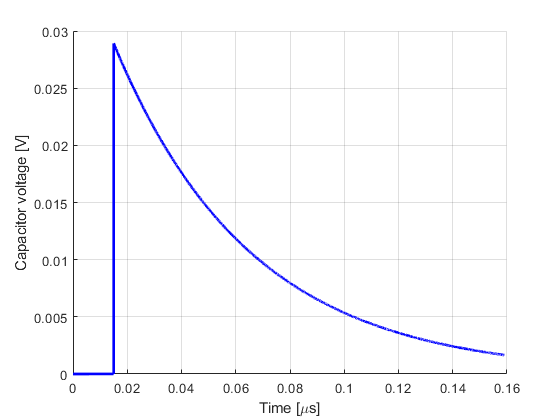}
    \caption{Time evolution of the capacitor voltage in the RC circuit with stationary ion between the plates.}
    \label{fig:validation_voltage_1}
\end{figure}
As a consequence of the potential difference induced between the capacitor electrodes, a counterclockwise current flows through the external circuit at each time step, as illustrated in Fig.~\ref{fig:validation_current_1}, leading to the accumulation of negative charge on the electrode surface. When the potential distribution induced by the accumulated surface charge is superposed with the ion-induced contribution, the resulting potential distribution is given by
\begin{equation}
\label{eq:potential_left_right_3}
\phi(x) =
\begin{cases}
\dfrac{q_\text{gap}}{\epsilon_0}\left(L - x_0\right)+\dfrac{q_\text{top}}{\epsilon_0}\left(L - x\right) & \text{for } x < x_0, \\[3ex]
\dfrac{q_\text{gap}}{\epsilon_0}\left(L - x\right)+\dfrac{q_\text{top}}{\epsilon_0}\left(L - x\right)  & \text{for } x > x_0.
\end{cases}
\end{equation}
Here, \(q_{\text{top}}\) represents the accumulated surface charge density on the circuit-side electrode.
Consequently, following ion injection, the spatial distribution of the electric potential in the capacitor gap exhibits the profile shown in Fig.~\ref{fig:validation_potential_1}. Furthermore, the temporal evolution of the potential difference between the two electrodes is presented in Fig.~\ref{fig:validation_voltage_1}.

%To investigate the difference between the strict form and the weak form, KVL is applied to the circuit shown in %Fig.~\ref{fig:validation_circuit_1} using BDF2 scheme, yielding
%\begin{equation}
%\label{eq:validation_circ}
%R_1\frac{3Q^n-4Q^{n-1}+Q^{n-2}}{2\Delta t}+\frac{Q^n}{C_1} = 0.
%\end{equation}
%In the strict form, Eq.~\eqref{eq:validation_circ} is reformulated using the capacitor  gap potential \(\phi^n\) evaluated at the n-th time step, resulting in 
%\begin{equation}
%\label{eq:validation_circ}
%R_1\frac{3Q^n-4Q^{n-1}+Q^{n-2}}{2\Delta t}+\phi^n = 0.
%\end{equation}
%Here, \(\phi^n\) denotes the electric potential that incorporates both the influence of the plasma domain and external circuit over the time interval \((n-1)\cdot \Delta t < t \le n\cdot\Delta t\). Similarly, \(Q^n\) denotes represents the total accumulated change in the circuit charge from the initial time up to \(t=n\cdot\Delta t\), including the contribution arising from plasma domain during the time interval \((n-1)\cdot \Delta t < t \le n\cdot\Delta t\). 

\subsection{Dynamic ion injection}

\begin{figure}[p]
    \centering
    \includegraphics[width=0.75\linewidth]{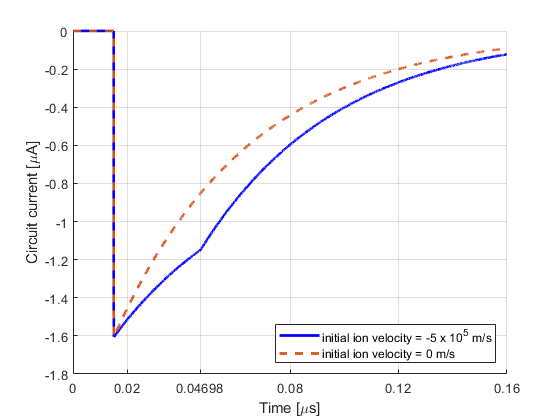}
    \caption{Time evolution of the circuit current in the RC circuit with an ion initially injected between the plates with a velocity of \(v_x =-5\times10^5\,\text{m/s}\).}
    \label{fig:validation_current_2}
\end{figure}

The second case considered involves a single ion generated at the same location and under the same simulation conditions as in the first case, but assigned an initial velocity of \(v_x = -5\times10^5\,\text{m/s}\), such that it is incident on the plate connected to the external circuit. As the ion approaches the electrode, the ion-induced potential increases according to Eq.~\eqref{eq:potential_left_right_2}, thus leading to an increase in the resulting circuit current flowing in the counterclockwise direction, compared to the stationary-ion case shown in Fig.~\ref{fig:validation_current_2}.
\begin{figure}[p]
    \centering
    \includegraphics[width=0.75\linewidth]{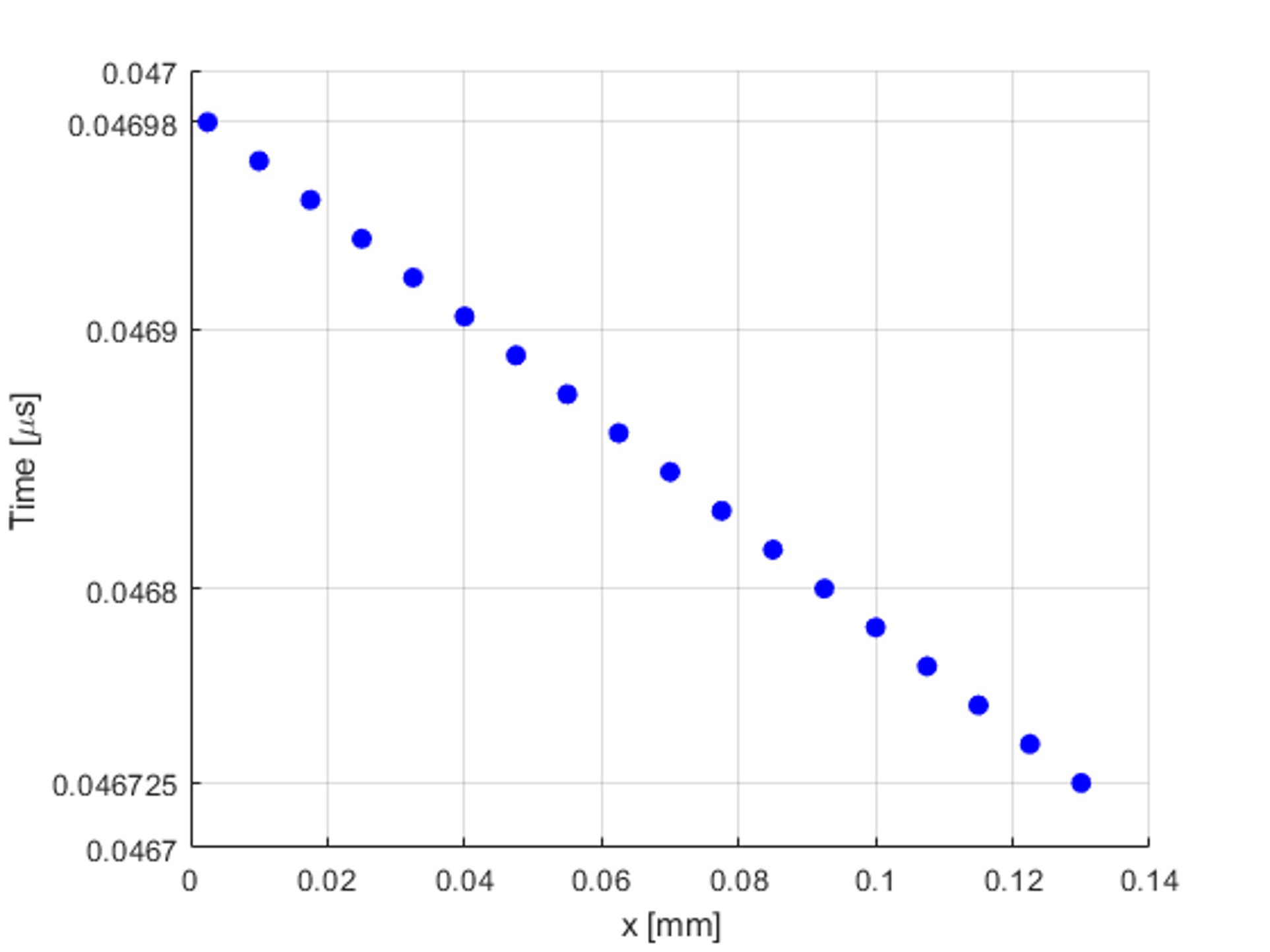}
    \caption{Time evolution of the ion trajectory after entering the first grid cell, where \(x=0\) corresponds to the electrode connected to the external circuit.}
    \label{fig:validation_trace}
\end{figure}
\begin{figure}[p]
    \centering
    \includegraphics[width=0.75\linewidth]{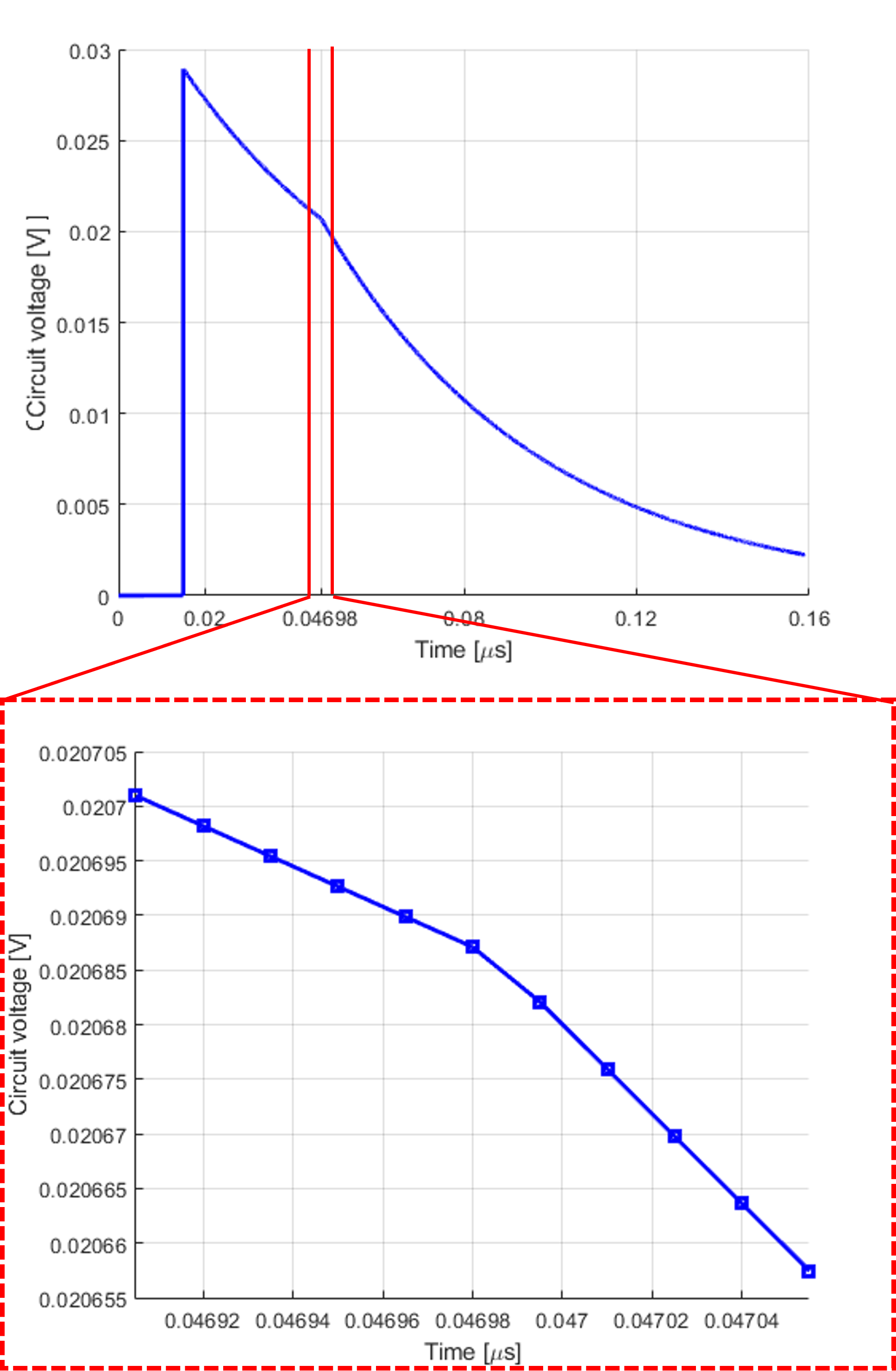}
    \caption{Time evolution of the capacitor voltage in the RC circuit with an ion initially injected between the plates with a velocity of  \(v_x =-5\times10^5\,\text{m/s}\).}
    \label{fig:validation_voltage_2}
\end{figure}
Under this initial condition, the ion reaches the top electrode just after \(t=47.98\)\SI{}{\micro\second}, as shown in Fig.~\ref{fig:validation_trace}. The potential distribution immediately before and after the ion enters the top electrode is continuous. Immediately before impact, the ion remains in the gap and approaches the electrode such that \(x_0\to 0\) in Eq.~\eqref{eq:potential_left_right_3}
Accordingly, the potential at the electrode surface is obtained as

\begin{equation}
\lim_{x_0\to 0^+}\phi(0) = \frac{q_\text{top} + q_\text{gap}}{\epsilon_0}L.
\end{equation}
If the ion is assumed to be fully absorbed by the electrode, the left-hand limit of the potential at the electrode surface yields the same expression, From the 
\begin{equation}
\lim_{x_0\to 0^-}\phi(0) = \frac{q_\text{top} + q_\text{gap}}{\epsilon_0}L.
\end{equation}
Therefore, the electric potential remains continuous across the electrode surface during ion absorption.

 As shown in Fig.~\ref{fig:validation_voltage_2}, the capacitor voltage evolution demonstrates that the continuity discussed above is well maintained in the simulation. This numerical continuity is achieved under the condition that, for $q_\text{gap}$ located between the 0th and 1st nodes, the charge densities at the first interior node and the boundary node are respectively interpolated as
\begin{equation}
\rho_1 = w_1\frac{q_\text{gap}}{\Delta x} \quad \text{and} \quad \rho_0 = w_0\frac{2 q_\text{gap}}{\Delta x},
\end{equation}
where
\begin{equation*}
w_0 = \dfrac{\Delta x - x_0}{\Delta x} \quad \text{and} \quad w_1 = \dfrac{x_0}{\Delta x} 
\end{equation*}
are the linear CIC weight coefficients.
This specific formulation implies that no charge is lost outside the physical boundary during the CIC allocation of the surface charge density.

\section{Comparison of weak and strict coupling modes}
\label{App_comp_strict_weak}
\begin{figure}[p]
    \centering
    \includegraphics[width=0.8\linewidth]{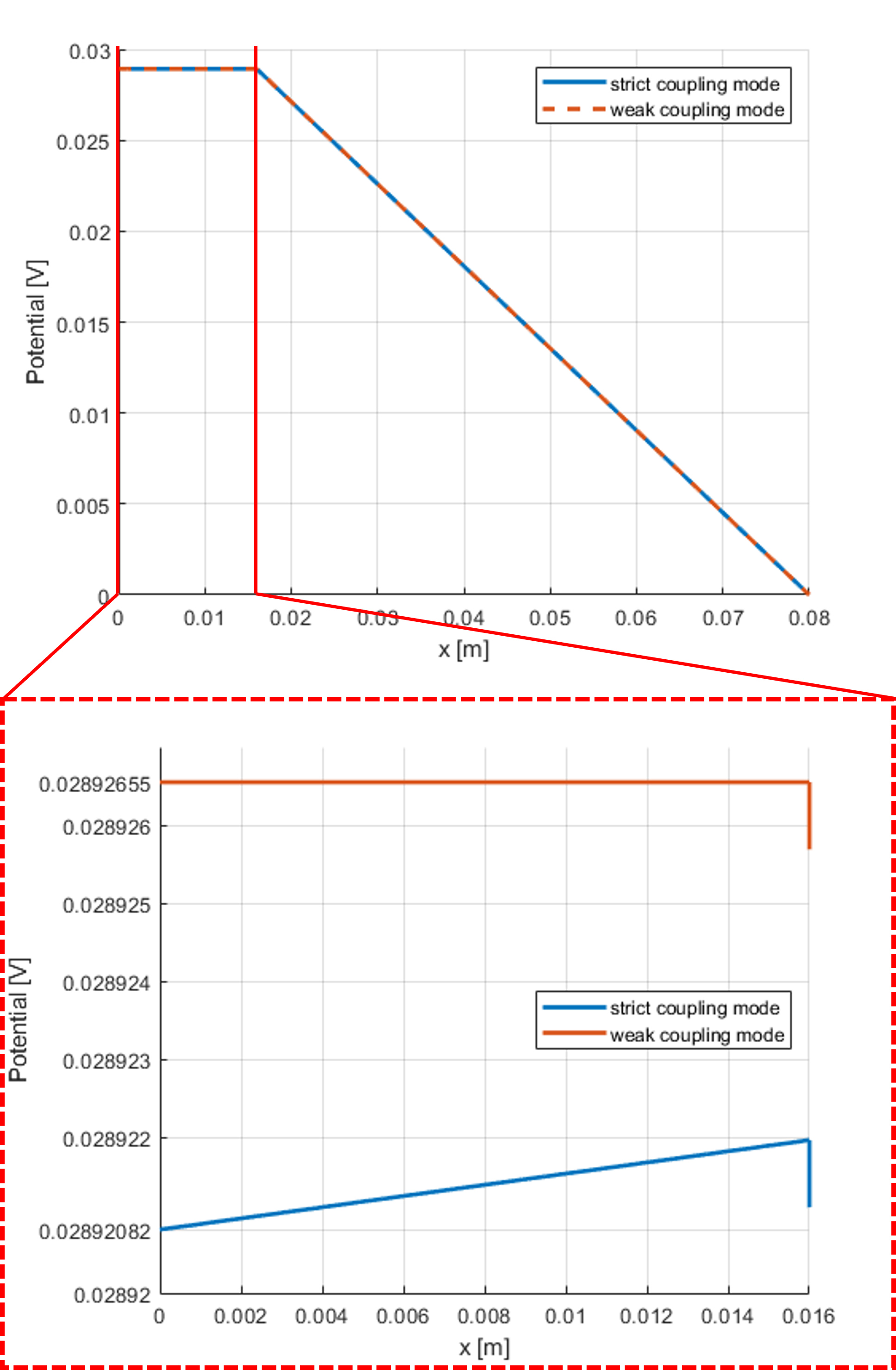}
    \caption{Potential distribution comparison between weak and strict coupling modes at the time step of stationary-ion injection in the capacitor gap.}
    \label{fig:validation_potential_3}
\end{figure}

In this appendix, the numerical behavior of the weak coupling mode and the strict coupling mode is analyzed using the RC circuit validation model presented in Appendix~\ref{App_validation}. In the strict coupling mode, the circuit equations and the plasma equations are implicitly coupled such that plasma events occurring at time step \(n\) are immediately reflected in the circuit charge \(Q^n_\text{circ}\) computed at the same time step. In contrast, in the weak coupling mode, \(Q^n_\text{circ}\) is determined independently of the plasma state at that time step, The influence of the plasma event manifests in \(Q^{n+1}_\text{circ}\) only after passing through the updated potential \(\phi^n\).

\begin{figure}[t]
    \centering
    \includegraphics[width=0.8\linewidth]{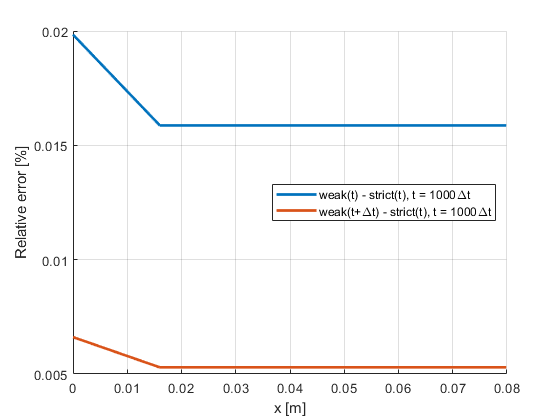}
    \caption{Relative errors in the voltage distribution of the weak coupling mode with respect to the strict coupling mode. The errors are evaluated at the same time step, and with the weak mode delayed by one time step.}
    \label{fig:validation_error}
\end{figure}

This behavior is observed in the simulation results, as illustrated in Fig.~\ref{fig:validation_potential_3}. When a stationary ion appears at \(x=0.016\,\text{m}\) at \(t=1000\cdot\Delta t\), the effect of the ion at that time step is not taken into account in the weak coupling mode, and the independently computed circuit charge remains zero, i.e., \(Q^n_\text{circ}=0\).
Consequently, the potential distribution is determined solely by the ion located in the capacitor gap, and the simulation result coincides with the analytical expression given in Eq.~\eqref{eq:potential_left_right_2}.

In the strict coupling mode, by contrast, the effect of the ion at the same time step is fully taken into account, and the circuit charge \(Q^n_\text{circ}\) and the electrode potential \(\phi^n(0)\) are implicitly determined so as to satisfy 
\begin{equation}
R_1 \dfrac{3Q^n_\text{circ}-4Q^{n-1}_\text{circ}+Q^{n-2}_\text{circ}}{2\Delta t} +\phi^n(0) = 0.
\end{equation}
For the present simulation, the resulting circuit charge is \(Q^n_\text{circ} =-1.606712\times10^{-17}\,\text{C}\). Since, under this configuration, the charge flowing into the top plate consists solely of the circuit current, the surface charge density on the top electrode can be expressed as
\begin{equation}
q_\text{top} = \dfrac{Q^n_\text{circ}}{A},
\end{equation}
where \(A\) denotes the area of the capacitor plate. Consequently, the blue curve in Fig.~\ref{fig:validation_potential_3} follows the analytical form given by Eq.~\eqref{eq:potential_left_right_3}

As shown in Fig.~\ref{fig:validation_potential_3}, the difference between the weak and strict coupling modes has a limited impact on the overall system behavior. In particular, as illustrated in Fig.~\ref{fig:validation_error}, the discrepancy can be interpreted as a simple one-time-step lag of the weak coupling mode relative to the strict coupling mode, while the spatial shape of the voltage distribution remains essentially unchanged.

\clearpage

\bibliographystyle{elsarticle-num} 
\bibliography{reference}

%% else use the following coding to input the bibitems directly in the
%% TeX file.

%% Refer following link for more details about bibliography and citations.
%% https://en.wikibooks.org/wiki/LaTeX/Bibliography_Management

\end{document}